\documentclass[aps,prb,twocolumn]{revtex4}
\usepackage{graphics,epsfig,amsfonts,amssymb,amsmath}
\begin{document}

\title{Tight-binding model for iron pnictides}
\author{M.J. Calder\'on$^{1}$}
\author{B. Valenzuela$^{1,2}$}
\author{E. Bascones$^{1}$}
\affiliation{$^1$Instituto de Ciencia de Materiales de Madrid,
ICMM-CSIC, Cantoblanco, E-28049 Madrid (Spain).\\
$^2$Departamento de la Materia Condensada, Universidad Aut\'onoma de Madrid,
Cantoblanco, E-28049 Madrid (Spain).
}
%\maketitle
\email{calderon@icmm.csic.es,belenv@icmm.csic.es,leni@icmm.csic.es}
\date{\today}
\begin{abstract}
We propose a five-band tight-binding model for the Fe-As layers of iron
pnictides with
the hopping amplitudes 
calculated within the Slater-Koster framework. The band structure found in 
DFT, including the orbital content of the bands, 
is well reproduced using only four fitting parameters to determine all the hopping 
amplitudes.
The model 
allows to study the changes in the electronic structure caused by a
modification of the angle $\alpha$ formed by the Fe-As bonds and the Fe-plane
and recovers the phenomenology 
previously 
discussed in the literature. We  also find that changes in $\alpha$ modify the shape and orbital content 
of the Fermi surface sheets.    
\end{abstract}
\maketitle
%\pacs{}
\section{Introduction}
\begin{figure}
\leavevmode
\includegraphics[clip,width=0.47\textwidth]{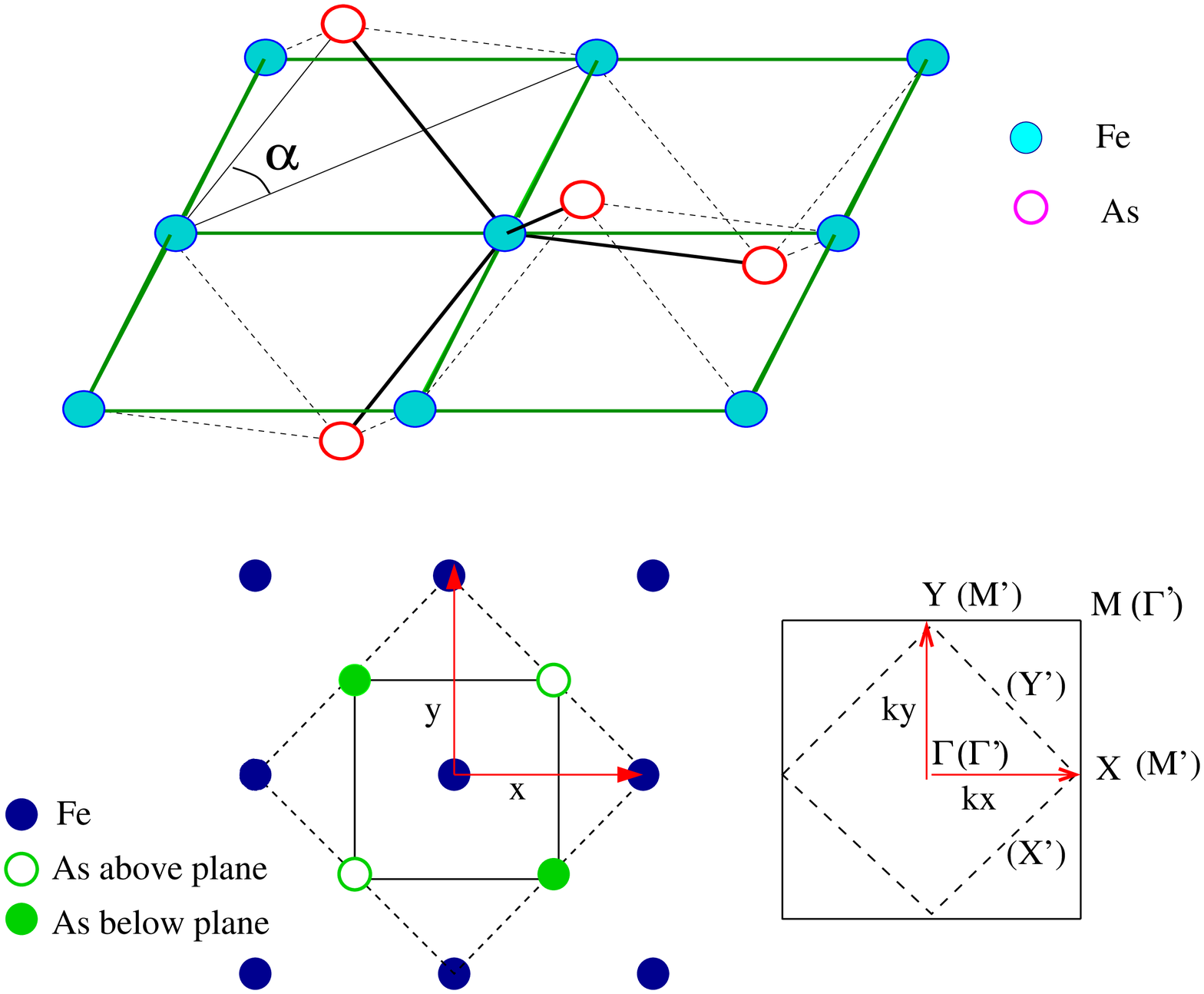}
\caption{
Top figure: Sketch of the lattice structure. 
Fe-As bonds form an angle $\alpha$ with
the Fe-plane which changes among compounds, with doping and with pressure.
Bottom figure: On the left, in a top view of the Fe-As layer, 
the real (extended) and the Fe unit cells are shown in dashed and solid lines
respectively. The X and Y axis of the Fe unit cell, used in the paper and, 
shown with arrows, are directed along the Fe bonds. On the right the
experimental (folded) Brillouin zone is shown with dashed lines. Its symmetry
points are denoted with primed letters.  The extended Brillouin zone, used in
the paper, is delimited by solid lines. It is double-sized and rotated 45 
degrees with respect to the experimental Brillouine zone. Bands and Fermi
pockets at $\Gamma$ and $M$ in the extended Brillouin zone and discussed
through the text will appear experimentally at $\Gamma'$. 
\label{fig:lattice}}
\end{figure}

Since the discovery of high temperature superconductivity in iron pnictides\cite{kamihara06,kamihara08} a
lot of attention has been devoted to their 
understanding. 
Iron pnictides are layered materials with 
arsenic (or another pnictogen) atoms 
at the center of the Fe plaquettes, out of plane and arranged upwards and
downwards in a checkerboard form (see Fig.~\ref{fig:lattice}) in  tetrahedral configuration. 
Fe-As bonds form an angle $\alpha$ with the Fe-plane, called in the following
Fe-As or iron-pnictogen angle, which differs among
compounds\cite{nomura09,ogino09} and depends on
doping\cite{rotter08,zhao08,nomura09}  
or applied pressure.\cite{kimber09} A possible connection between the 
value of $\alpha$, 
the critical temperature  and electronic properties has been
discussed by several authors.\cite{Lee08,mcqueen08,zhao08}  Recently Kuroki {\it et al}\cite{kuroki09-2}
have proposed that the pnictogen height above the Fe plane 
is the key factor that
determines both $T_c$ and the form of the superconducting gap.

From a DFT point of view 
iron superconductors have multiband character mostly due to Fe d-orbitals.\cite{lebegue07,singh08,mazin08-2}
In the iron (unfolded) Brillouin zone\cite{mazin08} the Fermi surface consists 
of electron pockets at the $X$ and $Y$ points, two hole pockets in $\Gamma$ 
and a hole pocket at $M$, in reasonable agreement with de Haas van
Alphen\cite{coldea08} experiments in the non-magnetic state.
  Angle-resolved photoemission (ARPES) measurements
  give also evidence of Fermi pockets at these symmetry
  points.\cite{liu08-2,liu08-zhou,lu08,ding08-2,zabolotnyy08}
Interband scattering between electron and hole pockets has
been proposed as a mechanism for superconductivity.\cite{mazin08,yao09} 
In this context the importance of nesting for superconductivity and magnetism
is discussed\cite{raghu08,korshunov08}. More recently, the relevance of the anisotropic
orbital weight of each Fermi pocket in determining the symmetry of the
superconducting order parameter has been emphasized\cite{maier09,zhai09,kuroki09-2}.   
Electron pockets at $X$ and $Y$ have respectively 
$yz/xy$ and $zx/xy$ origin while the hole pockets in $\Gamma$ arise 
from $zx$ and $yz$ orbitals\cite{boeri08}. Due to
closeness of two hole bands and different dependence of their energy
on $\alpha$  the orbital character of the pocket in $M$ switches
between $xy$ or $3z^2-r^2$ depending on the value of 
$\alpha$.\cite{vildosola08,lebegue09} Experimentally, the orbital content 
can be studied by
changing the polarization of the light used in ARPES.\cite{hsieh08,fink09,shimojima09,zhang09}

A good tight-binding model
is the basic building-block of any theoretical treatment in a lattice. 
Initial attemps tried to describe the iron pnictides using two\cite{raghu08,daghofer08,stanescu08} or 
three-orbital\cite{patricklee08} models.  Several proposals based on the symmetry
properties of $zx$ and $yz$ (and $xy$ as  third orbital) were put forward to
describe the bands close to the Fermi level. 
However crystal field splittings among the Fe d-orbitals are small compared
with the bandwidth resulting in strong hybridization of all d-orbitals.  
At present  it is believed that inclusion of all five Fe d-orbitals is 
necessary to obtain a good description of the properties of iron pnictides\cite{kuroki08,cvetkovic09,eschrig09}. 
The placement of As at the center of the plaquettes suggests that hopping
between Fe atoms to second nearest neighbors cannot be disregarded.

In this paper we propose a five-orbital tight binding model to describe the
Fe-As layers with 
the hopping amplitudes 
calculated within the Slater-Koster framework.\cite{slater54}  Compared to DFT tight-binding
fits, the procedure presented here greatly reduces
the number of fitting parameters necessary to calculate the bands
and allows to study changes in the iron-pnictogen angle $\alpha$. We
show that the bands close to the Fermi level can be described giving all the
hopping amplitudes in terms of just four parameters. The agreement between
our results and DFT predictions extends to the orbital weight of each band. We
 also reproduce the 
switch in orbital character of the hole pocket in M, when $\alpha$ varies.
Furthermore, we predict that
changes in $\alpha$ can 
induce modifications in the shape and orbital
content of the Fermi pockets, including the disappearance of the hole 
pockets in $\Gamma$ 
when the tetrahedron is elongated. Within the present theoretical 
understanding these results have strong implications in the superconducting 
and magnetic properties of these compounds.\cite{maier09,kuroki09-2,zhai09}

\section{Tight-binding model}
We construct a tight-binding model to describe the band structure of the FeAs
layers  including the five Fe d-orbitals.  Arsenic atoms only enter in the model indirectly via 
the Fe-Fe hopping amplitudes. Indirect hopping via arsenic is treated to 
second order in perturbation theory. Direct hopping between Fe atoms is also
included. Hopping is restricted to first and 
second nearest Fe neighbors. Both the mathematical form of the Hamiltonian and the hopping
amplitudes are computed within the Slater-Koster formalism\cite{slater54}. We take 
$X$ and $Y$  directed 
along the Fe-bonds  (see Fig.\ref{fig:lattice}). The same axis convention applies for the  
orbitals, i.e. $x^2-y^2$ orbital lobes are directed along the Fe-Fe bonds. Under these assumptions the Hamiltonian is given by: 
\begin{widetext}
\begin{eqnarray}
H=\sum_{m,n,\sigma}\left (\sum_{\gamma}\left [
\epsilon_\gamma d^\dagger_{m,n;\gamma,\sigma}d_{m,n;\gamma,\sigma}
  +\sum_{s_x=\pm 1} t^x_{\gamma,\gamma}
    d^\dagger_{m+s_x,n;\gamma,\sigma}d_{m,n;\gamma,\sigma} + 
\sum_{s_y=\pm 1} t^y_{\gamma,\gamma}
    d^\dagger_{m,n+s_y;\gamma,\sigma}d_{m,n;\gamma,\sigma}
\right. \right. \nonumber \\ \left. \left. +  \sum_{s_x,s_y=\pm 1}\tilde
  t_{\gamma,\gamma} d^\dagger_{m+s_x,n+s_y;\gamma,\sigma}d_{m,n;\gamma,\sigma}    
\right ]
+\sum_{[\gamma\neq \beta]}t_{\gamma,\beta} 
\left [\sum_{s_x=\pm 1}d^\dagger_{m+s_x,n;\beta,\sigma}d_{m,n;\gamma,\sigma}-
    \sum_{s_y=\pm 1}d^\dagger_{m,n+s_y;\beta,\sigma}d_{m,n;\gamma,\sigma}\right
  ]\right. \nonumber \\ \left.
+\sum_{\langle \gamma \neq \beta \rangle}\sum_{s_x,s_y=\pm 1}s_xs_y\tilde t_{\gamma,\beta}
 d^\dagger_{m+s_x,n+s_y;\beta,\sigma}d_{m,n;\gamma,\sigma}
+(-1)^{m+n}
\left [\sum_{(\gamma\neq \beta)}\sum_{s_x=\pm 1}s_xt^x_{\gamma,\beta}d^\dagger_{m+s_x,n;\beta,\sigma}d_{m,n;\gamma,\sigma}
 \right. \right.  \nonumber \\ \left. \left. + \sum_{((\gamma\neq \beta))}\sum_{s_y=\pm 1}s_yt^y_{\gamma,\beta}d^\dagger_{m,n+s_y;\beta,\sigma}d_{m,n;\gamma,\sigma}
 + \sum_{(\gamma\neq \beta)}\sum_{s_x,s_y=\pm 1}s_x\tilde t_{\gamma,\beta}d^\dagger_{m+s_x,n+s_y;\beta,\sigma}d_{m,n;\gamma,\sigma}
\right. \right. \nonumber \\ \left. \left. 
+ \sum_{((\gamma\neq \beta))}\sum_{s_x,s_y=\pm 1}s_y\tilde t_{\gamma,\beta}d^\dagger_{m+s_x,n+s_y;\beta,\sigma}d_{m,n;\gamma,\sigma}
\right]
  \right)-\mu
\label{tightbindingreal}
\end{eqnarray}
\end{widetext}
Here $m,n$ refer to lattice sites, $\gamma,\beta$ are the orbital indices,
$\sigma$ the spin and $\mu$ the chemical potential. 
Only the first sum runs
through all the orbitals. 
Brackets and parentheses restrict the orbitals to which the 
other sums apply. In
particular, $\langle \gamma\neq \beta \rangle$ is restricted to  the
pairs $\{\gamma,\beta\}=\{yz,zx\}$, 
$\{xy,3z^2-r^2\}$, 
$[\gamma\neq\beta]$ to the pair
$\{\gamma,\beta\}=\{3z^2-r^2,x^2-y^2\}$, $(\gamma\neq \beta)$ to  the pairs
$\{\gamma,\beta\}=\{yz,3z^2-r^2\},\{yz,x^2-y^2\},\{zx,xy\}$, and
$((\gamma\neq \beta))$ to  the pairs
$\{\gamma,\beta\}=\{yz,xy\},\{zx,3z^2-r^2\},\{zx,x^2-y^2\}$. 
$\epsilon_\gamma$ are the on-site energies of the $d$ orbitals. Due to the
degeneracy of $yz$ and $zx$, $\epsilon_{yz}=\epsilon_{zx}$. 
From the orbital symmetry it follows
$t^x_{\gamma,\gamma}=t^y_{\gamma,\gamma}$ for
$\gamma=xy,3z^2-r^2,x^2-y^2$. Second nearest neighbor hopping parameters 
$\tilde t_{\gamma,\beta}$ where
$\gamma=xz,yz$ and $\beta=xy,3z^2-r^2,x^2-y^2$ change sign when $\gamma$ and
$\beta$ orbitals are exchanged. In any other case 
$\tilde t_{\gamma,\beta}=\tilde t_{\beta,\gamma}$ and
$t^a_{\gamma,\beta}=t^a_{\beta,\gamma}$, with $a=x,y$. 
Other equalities brought by
the symmetry are: 
\begin{eqnarray}
t^x_{zx,zx}& = & t^y_{yz,yz},\nonumber \\ 
t^y_{zx,zx}& = & t^x_{yz,yz}, \nonumber \\
t^x_{zx,xy}& = & t^y_{yz,xy}, \nonumber \\
t^y_{zx,3z^2-r^2}& = & t^x_{yz,3z^2-r^2} \nonumber \\
t^y_{zx,x^2-y^2} & = & -t^x_{yz,x^2-y^2} \nonumber \\
\tilde t_{yz,xy} & = & \tilde t_{zx,xy} \nonumber \\
\tilde t_{yz,3z^2-r^2} & = & \tilde t_{zx,3z^2-r^2} \nonumber \\
\tilde t_{yz,x^2-y^2} & = & -\tilde t_{zx,x^2-y^2}
\label{equalities}
\end{eqnarray}

\begin{figure}
\leavevmode
\includegraphics[clip,width=0.52\textwidth]{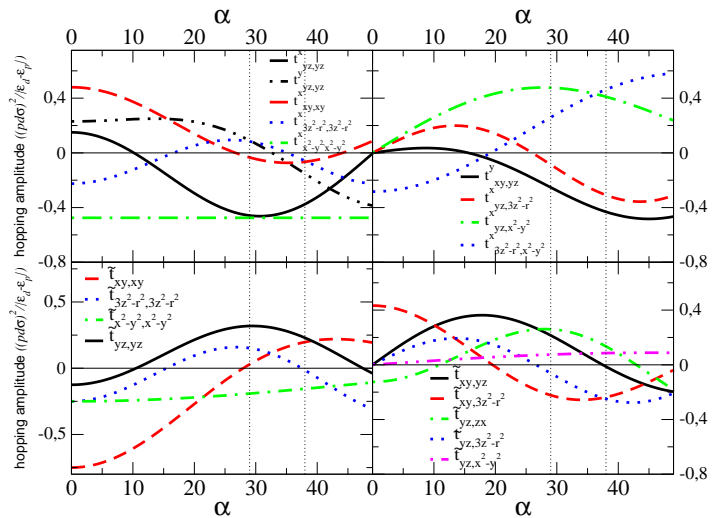}
\caption{  
Dependence of the hopping
amplitudes on $\alpha$. Experimental values of $\alpha$ are between $29$ and
$38$ degrees. 
Top (bottom) graphs: first (second) nearest neighbor hopping amplitudes
corresponding to 
$pd\pi$=-0.5, $(dd\sigma)_1=-0.6$,
$(dd\pi)_1=0.48$ and $(dd\delta)_1=-0.1$. 
Direct Fe-hopping between second nearest neighbors via 
$(dd\sigma)_2$, $(dd\pi)_2$ and $(dd\delta)_2$ is neglected. 
All the energies
are in units of $(pd\sigma)^2/|\epsilon_d-\epsilon_p|$ except $pd\sigma$ and 
$pd\pi$ which are in units of $pd\sigma$. Here $\epsilon_p$ and $\epsilon_d$ are 
the on-site energies of the pnictogen-p and the Fe-d orbital (see Appendix I).
The same fitting parameters and energy units are used through all the text.  
}
\label{fig:parameters}
\end{figure}

The complex sign structure  of the hopping terms included in  the $s_x$ and $s_y$
factors arises from changes of 
sign in the orbital wave functions. 
The factor $(-1)^{m+n}$
in the terms which mix $yz,zx$ with $xy,3z^2-r^2,x^2-y^2$ reflect the doubling
of the unit cell due to the checkerboard alternance of the arsenic atoms
displaced up
and down from the center of the Fe-square plaquettes.
These terms
vanish when the arsenic atoms are in the Fe-planes (see Fig.~\ref{fig:parameters} and Appendix I). 
Due to the enlargement of the unit cell, in the reduced Brillouin zone  
 $-\frac{\pi}{2}< k'_x,k'_y <\frac{\pi}{2}$, the Hamiltonian is a $10 \times
 10$ matrix. As discussed in the context of three and four band
 models\cite{patricklee08,yu09} and in Appendix II, 
it is possible to work in an unfolded Brillouin zone $ -\pi< k_x,k_y <\pi$
where ${\bf k}={\bf k'}$ for orbitals $yz$ and $zx$ and  ${\bf k}={\bf k'+Q}$
in the case of  $xy, 3z^2-r^2$, and $x^2-y^2$. In this unfolded Brillouin zone,  
the system is described by a
five-band Hamiltonian  $H_{5\times 5}({\bf k})$. The relation between the
  unfolded and the reduced Brillouin zones is displayed in Fig.~\ref{fig:lattice}.

In  previous five-band Hamiltonians,
\cite{kuroki08,graser09} 
the hopping amplitudes $t^x_{\gamma,\beta}$, $t^y_{\gamma,\beta}$ and
$\tilde t_{\gamma,\beta}$  were determined from a fitting to DFT
bands. In contrast, here they are calculated within the Slater-Koster
framework\cite{slater54}. 
This method had been used before in two and three-band 
models for iron pnictides.\cite{daghofer08,calderon09,moreo09}   It
involves a small number of fitting parameters as all the hopping
terms depend on a few disposable constants, the Fe-As and Fe-Fe orbital
overlap integrals. 
The final expressions for the
hopping amplitudes are given in Appendix I. 
Indirect hopping via
As induces a dependence of the hopping amplitudes on the angle $\alpha$ formed
by Fe-As bonds and the Fe plane. 
This dependence is shown in Fig.~\ref{fig:parameters}.  
In the range of experimental interest of $\alpha$ ($29^o-38^o$) 
a strong variation of the hoppings is seen indicating important 
implications for any proposed model -either in the weak\cite{mazin08,raghu08,korshunov08,chubukov08} or strong coupling 
limit\cite{yildirim08,daghofer08,si08,haule08}- to describe these compounds.

In Fig.~\ref{fig:parameters} and in the rest of the paper 
the values of the overlap integrals and crystal field parameters 
have been chosen to  reproduce the main features of the band structure of 
LaFeAsO when $\alpha$ equals
the angle measured experimentally\cite{delacruz08} in this compound
$\alpha^{LaFeAsO}=33.2^o$. 
While the expressions for the hopping amplitudes
given in Appendix I include both Fe-Fe overlap integrals up to second nearest
neighbors, we have found that the bandstructure is well described including
only Fe-As and Fe-Fe overlap to nearest neighbors and neglecting Fe-Fe direct overlap to
second nearest neighbors. With such a choice, hopping between Fe atoms to 
next nearest neighbors is completely mediated by As, while hopping to first
nearest neighbors has contributions from both direct hopping between Fe atoms
and indirect hopping via As. 

The same values  for
the overlap integrals and crystal field splittings are
used in all the figures through the paper. 
In section IV we analyze the effect of changing $\alpha$ on the band
structure, using in the analysis values of $\alpha$ which have been found
experimentally in several pnictides. However,
we caution on the application of the results obtained here on the
angle dependence  to compare different compounds. Substitution of arsenic atoms by
P or a change in the lattice constant could modify to some extent the 
values of the integral overlaps.

\section{Band structure for $\alpha^{LaFeAsO}$}

\begin{figure}
\leavevmode
\includegraphics[clip,width=0.48\textwidth]{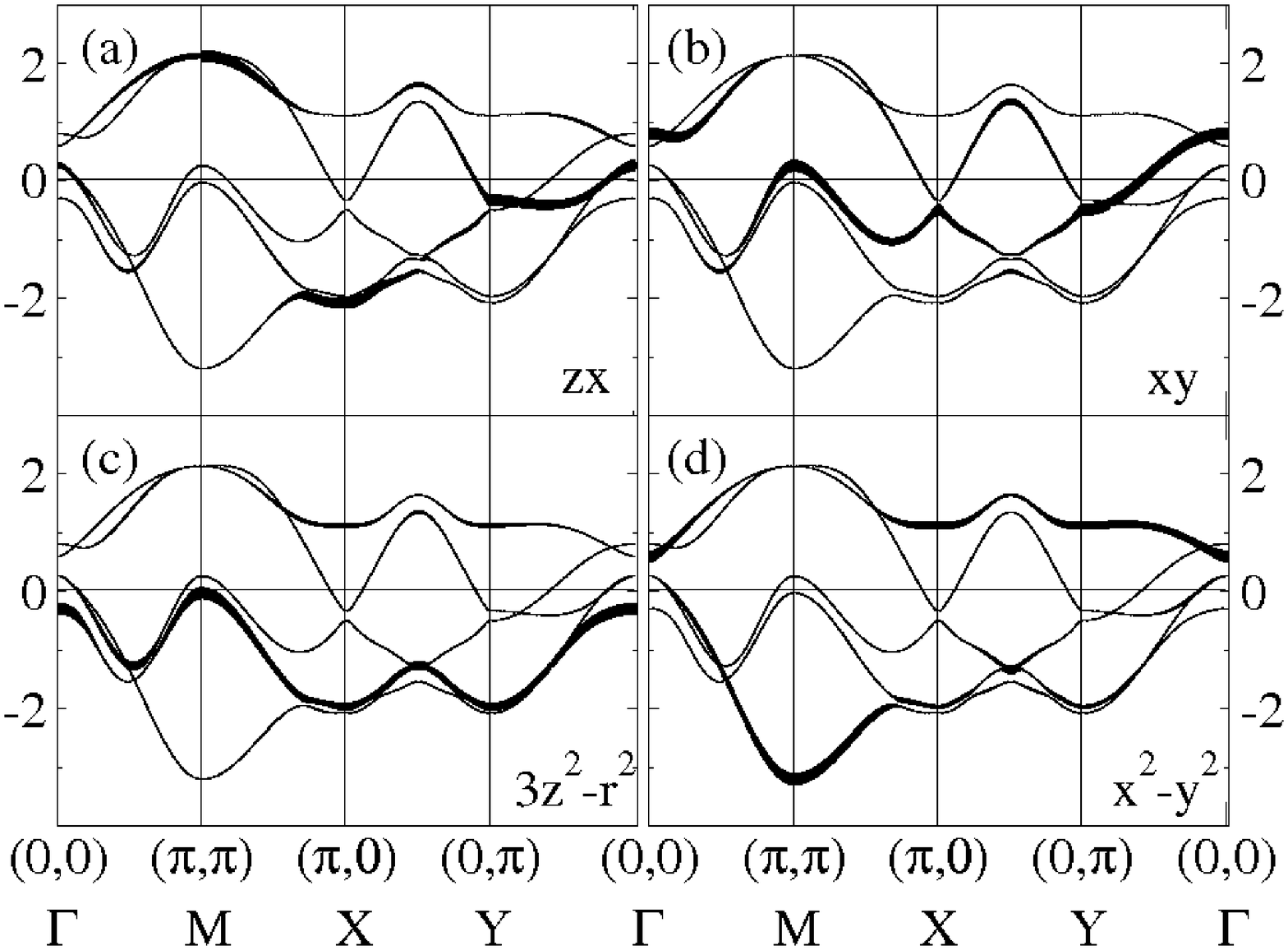}
\caption{
Band structure in the unfolded Brillouin zone obtained from the tight-binding Hamiltonian (1) with the
hopping amplitudes computed within the Slater-Koster framework as described 
in Appendix I. Values used for the overlap integrals are given in Fig.~\ref{fig:parameters}, 
$\alpha^{LaFeAsO} =33.2^o$, as found experimentally in LaFeAsO. For
onsite-energy values see text. From (a) to (d) the width of each band-line is
proportional to its $zx$, $xy$,
$3z^2-r^2$ and $x^2-y^2$ weight.  
}
 \label{figlafeasO}
\end{figure}

Fig.~\ref{figlafeasO} shows the band structure obtained from (1)  for 
$\alpha=33.2^o$,  the 
experimental Fe-As angle in LaFeAsO, and the overlap integrals given
in Fig.~\ref{fig:parameters}. 
For the crystal field splitting we take $\epsilon_{xy}=0.02$,
$\epsilon_{zx,yz}=0$, which defines the zero of energy, 
$\epsilon_{3z^2-r^2}=-0.55$ and
$\epsilon_{x^2-y^2}=-0.6$, in units of $(pd\sigma)^2/|\epsilon_d-\epsilon_p|$. 
The order of the onsite energies taken here has
been discussed extensively in the literature and the values chosen are similar
to those  used previously by other authors\cite{kuroki08,haule09,yu09,maier09,eschrig09}. 
The Fermi level corresponds to
filling the bands with six electrons (including spins), as found in undoped 
pnictides or compensated FeAs layers.  All the figures are shown in the Fe or
unfolded Brillouin zone. 
Bands in Fig.~\ref{figlafeasO} have a strong resemblance with those
obtained from LDA calculations,  once they are represented in the unfolded
  Brillouin zone. Pockets at the Fermi level include: two hole
pockets at $\Gamma=(0,0)$, a hole pocket in $M=(\pm \pi,\pm \pi)$ and 
electron pockets in $X=(\pm \pi,0)$ and $Y=(0,\pm \pi)$. 
The resulting Fermi surface is plotted in Fig.~\ref{figtodasfs}.
The two-hole pockets in $\Gamma$
originate in two hole-bands degenerate at the top due
to the degeneracy of $zx$ and $yz$ orbitals. 
The so-called Dirac  point\cite{kuroki09-2,zhai09} which results from the
  crossing of $zx$ and $xy$ derived bands is located close to the Fermi level in
  the vicinity of $(0,\pi)$. From the expression of the Hamiltonian in
  Appendix II, it can be seen that these two orbitals do not mix in the
  $(0,0)-(\pi,0)$ direction.

The agreement between our results and LDA calculations also 
extends to their orbital character. From top to bottom in 
Fig.~\ref{figlafeasO} we show the energy bands weighted by  their $zx$, $xy$, 
$3z^2-r^2$ and $x^2-y^2$ orbital content, which can be compared with the
results by Boeri {\it et al}.\cite{boeri08} The $yz$ weight is equivalent to the $zx$
weight if $X$ and $Y$ axis are  interchanged. The two hole-bands in $\Gamma$ which
cross the Fermi level have mostly $zx$ and $yz$ character, while some
$x^2-y^2$ weight can also be appreciated. The electron pockets at  
$(\pm \pi,0)$/$(0,\pm \pi)$ arise from $yz/zx$ and $xy$ orbitals. 
$3z^2-r^2$ contributes mostly to bands below the
Fermi level. The orbital content of the Fermi pockets is better
   seen in Fig.~\ref{figfsorbitalcontent}.

Around $(\pm \pi,\pm \pi)$
a $3z^2-r^2$ hole-band nearly touches the Fermi level, without crossing it.
This band is close to another
$xy$ hole-band which produces the pockets at
$(\pm \pi,\pm \pi)$. All these features are also present in the LDA bands. The 
pocket at $(\pi,\pi)$ has been a matter of discussion in the literature. 
Initially\cite{singh08}, it was proposed that there was a small three-dimensional 
pocket with $3z^2-r^2$ character which, in the reduced Brillouin zone 
appeared {\bf at $\Gamma'$}. 
This conclusion was reached using the relaxed lattice structure
and not the experimental one. 
It was later shown that using the experimental
lattice parameters, in particular the experimental Fe-As angle, the position
of the top of the $xy$ and  $3z^2-r^2$ hole-bands at  $\Gamma'$ in the
  reduced Brillouin zone switch and
the Fermi pocket has $xy$ character\cite{vildosola08}.
This apparent
disagreement between the results obtained with the relaxed and the
experimental lattices originates in a strong dependence of the band structure
on the Fe-As angle\cite{mazin08-2,boeri08,vildosola08,lebegue09}.  We show in Sec. IV that the tight binding proposed here
reproduces this angle-dependence of the band energies  for the pocket
  which appears at $(\pi,\pi)$. 

\begin{figure}
\leavevmode
\includegraphics[clip,width=0.67\textwidth,angle=90]{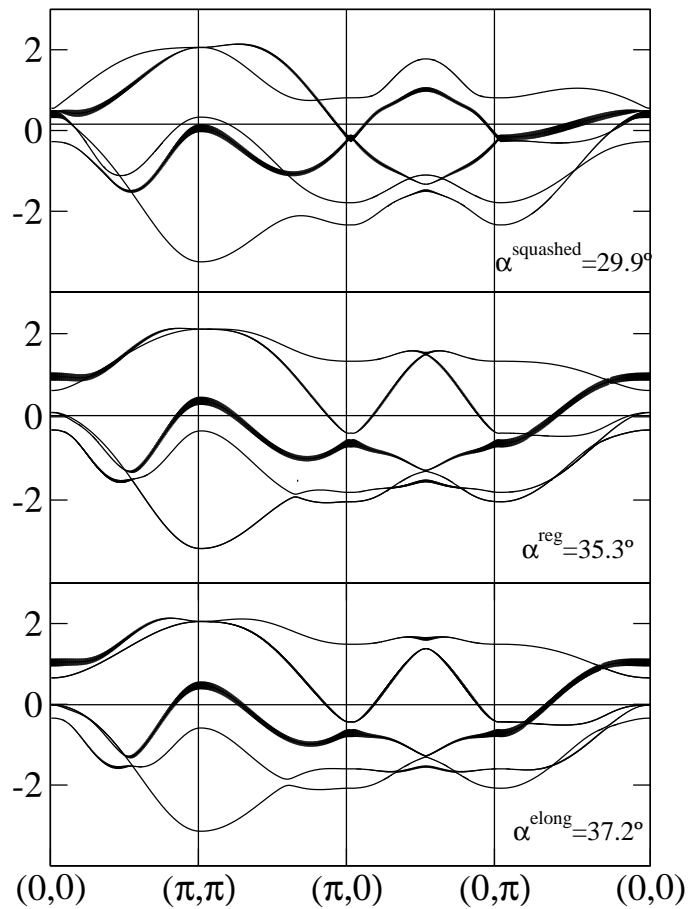}
\caption{
From top to bottom, energy bands corresponding to
$\alpha^{squashed}=29.9^o$ (as found in LaFePO), $\alpha^{reg}=35.3^o$ (regular
tetrahedron),
and $\alpha^{elong}=37.2^o$ (elongated tetrahedron).  The width of the curves is proportional to its $xy$ weight.} 
\label{figbandasangulo}
\end{figure}

The good agreement (shown in Fig.~\ref{figlafeasO}) is not restricted to
the energies closest to the Fermi level, but it is quite generic to all the
bands. The correspondence is more impressive having in mind that all the hopping
amplitudes are given in terms of just four free parameters.  We emphasize
that
  the orbital overlaps and crystal field values have not
  been optimized to fit the LDA bands of LaFeAsO, but just correspond to 
  the minimum set of parameters that reproduce the qualitative features 
  of the band structure using the expected orbital energy splitting.

As discussed above the value of the overlap integrals is expected to depend to
some extent on the lattice parameters and atomic radii. We have found
that the band structure is sensitive against small changes in
the fitting parameters.
Close to the Fermi level the largest variations appear in the relative
position of the top of the $xy$ and $3z^2-r^2$ hole bands in $M$ between them 
and with respect to those in $\Gamma$, and the energies of $xy$ and $zx$ bands
in $Y$.  
This 
behavior might be an indication of the  experimentally found strong 
sensitivity of these compounds 
to modifications in structural parameters.

\section{Fe-As Angle Dependence}

We now focus on the changes in the band structure produced by a modification
of the angle $\alpha$. We assume that all dependence enters via the hopping
amplitudes. The crystal field splitting of the Fe $d$ orbitals results from both
the As and Fe environment of each Fe atom. Modifications in $\alpha$ change
the electrostatic environment produced by arsenic atoms but not the one due to Fe
atoms. We assume that in the range of $\alpha$ values of interest 
the change in the crystal field parameters is small and we 
neglect the dependence of the onsite energies on
$\alpha$. As discussed in Appendix II to analyze the effect of a possible
change of crystal field with $\alpha$ is straightforward. 
In Fig.~\ref{figbandasangulo} we plot the energy bands corresponding to   
$\alpha^{squashed} =29.9^o$ (squashed tetrahedron),
$\alpha^{reg}=35.3^o$ (regular tetrahedron) and $\alpha^{elong}=37.2^o$ ( 
elongated tetrahedron) and the fitting parameters used in Fig.~3. 
The bandline width is proportional to the weight of the $xy$ orbital. 
The first
two  values of $\alpha$ used 
have been found in LaFePO, and in 
BaFe$_2$As$_2$ at optimal doping respectively. The FeAs$_4$ tetrahedron is 
slightly elongated in CaFeAsF.

\begin{figure}
\leavevmode
\includegraphics[clip,width=0.50\textwidth]{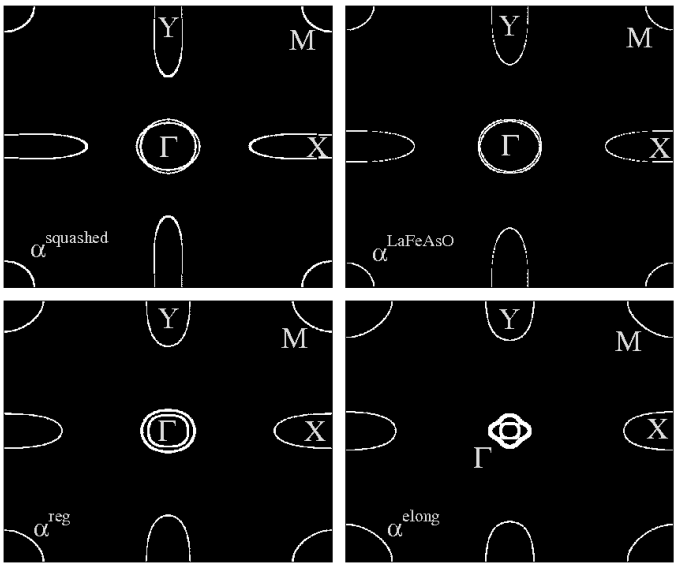}
\caption{
Fermi surface corresponding to
$\alpha^{squashed}=29.9^o$ (as found in LaFePO), $\alpha^{LaFeAsO}=33.2^o$ (as found in LaFeAsO),
$\alpha^{reg}=35.3^o$ (regular tetrahedron) and $\alpha^{elong}=37.2^o$ 
(elongated tetrahedron) with the same fitting parameters as in Fig.~3.}
 \label{figtodasfs}
\end{figure}

 As evident in Fig.~\ref{figbandasangulo}, even small modifications of the Fe-As angle have an impact on the band structure.
Around the Fermi level
$E_F$, the most clear change appears close to $(\pi,\pi)$. There are two 
hole bands with maximum close to $E_F$, with mostly $xy$ or $3z^2-r^2$ 
character. For $\alpha^{squashed}$ the $3z^2-r^2$ is higher in energy and crosses
the Fermi level. However the relative position of the two bands changes
as $\alpha$ increases and the hole pocket around  $(\pi,\pi)$ has $xy$
character in the other two cases considered here. The energy difference
between both bands increases as the tetrahedron is elongated. This behavior
was first obtained from LDA calculations. 
In our tight-binding model it is easy to understand the origin of 
this shift. At $(\pi,\pi)$ the energy of $xy$ and $3z^2-r^2$ orbitals 
is $E_\beta (\pi,\pi)=4t^x_\beta + \tilde t_\beta$ for $\beta=xy$, $3z^2-r^2$. 
The upwards shift in $E_{xy}(\pi,\pi)$ with increasing $\alpha$ is due to the 
increase of $xy$ second nearest neighbors $\tilde t_{xy}$, 
while the first nearest neighbors $t^x_{xy}$ 
remains almost constant (see Fig.~\ref{fig:parameters}). On the other hand both $t^x_{3z^2-r^2}$ 
and $\tilde t_{3z^2-r^2}$ decrease when $\alpha$ increases.

The dependence of the band structure on the  Fe-As angle is also seen at
$\Gamma$. The gap between the top of the hole bands which cross $E_F$ and the
$xy$ band at higher energies is strongly reduced with decreasing $\alpha$,
in part due to a change in $\tilde t_{xy,xy}$ (see Appendix II). This gap reduction is,
however, not only due to a decrease in energy of the $xy$ band. The top of the
$yz,zx$ hole bands shifts upwards as $\alpha$ is reduced. On the other hand,
elongation of the tetrahedron can lead to the disappearance of the hole pockets
at $\Gamma$. 

As $\alpha$ decreases, a transfer of $xy$ orbital weight from the third to the
second band can be appreciated in the $(\pi,0) \rightarrow (0,\pi)$
direction. This is accompanied by a shift 
of the Dirac point
towards $(0,\pi)$. Other changes in the band structure with $\alpha$ are
discussed in Appendix~\ref{app:hamk}.

Somehow weaker, but still observable is the
change in shape of the electron and hole pockets at $\Gamma$ and $X$ ($Y$). 
This feature is better observed in
Fig.~\ref{figtodasfs}. The electron pockets at $X$ and $Y$ are more elongated 
towards $\Gamma$ as $\alpha$ is reduced. The shape of hole pockets is
qualitatively modified as $\alpha$ increases. For the smallest angle, 
$\alpha^{squashed} =29.9^o$, the hole Fermi pockets at $\Gamma$ resemble two 
ellipses centered
at $\Gamma$ with axis
directed along $X$ and $Y$ directions. With increasing $\alpha$, 
For both $\alpha^{squashed} =29.9^o$ and 
$\alpha^{LaFeAsO}=33.2^o$ the
  two Fermi sheets are very close to each other and would be hardly
  distinguishable in ARPES or quantum oscillation experiments.  
With increasing $\alpha$, for  a value 
corresponding to a regular tetrahedron we find two concentric
pockets.
 Finally, when the tetrahedron
is elongated the inner hole has a square-like shape while the outer one has a
flower-like shape. 
 Similar Fermi surfaces have been found in an ab-initio study of the effect of
pressure in the 122 family\cite{opahle09}. 
Both circular-like and  square-like hole pockets at $\Gamma$ have been
reported from ARPES measurements\cite{liu08-2,liu08-zhou,lu08,ding08-2,zabolotnyy08,hsieh08,fink09,shimojima09,zhang09}.
We emphasize that we are working in the unfolded Brillouin zone. ARPES
experiments sample the folded Brillouin zone  where the pocket that we found at
$M$ would be also expected at $\Gamma$. Its relative size, compared to the
other two hole pockets in $\Gamma$, will depend on the Fe-As angle.

In our model, the
exact shape of the hole Fermi pockets found for a given angle 
can depend on the exact fitting parameters
used.
However, the change of the Fermi pockets shape with 
$\alpha$ is a robust feature.

\begin{widetext}

\begin{figure}
\includegraphics[clip,width=0.98\textwidth]{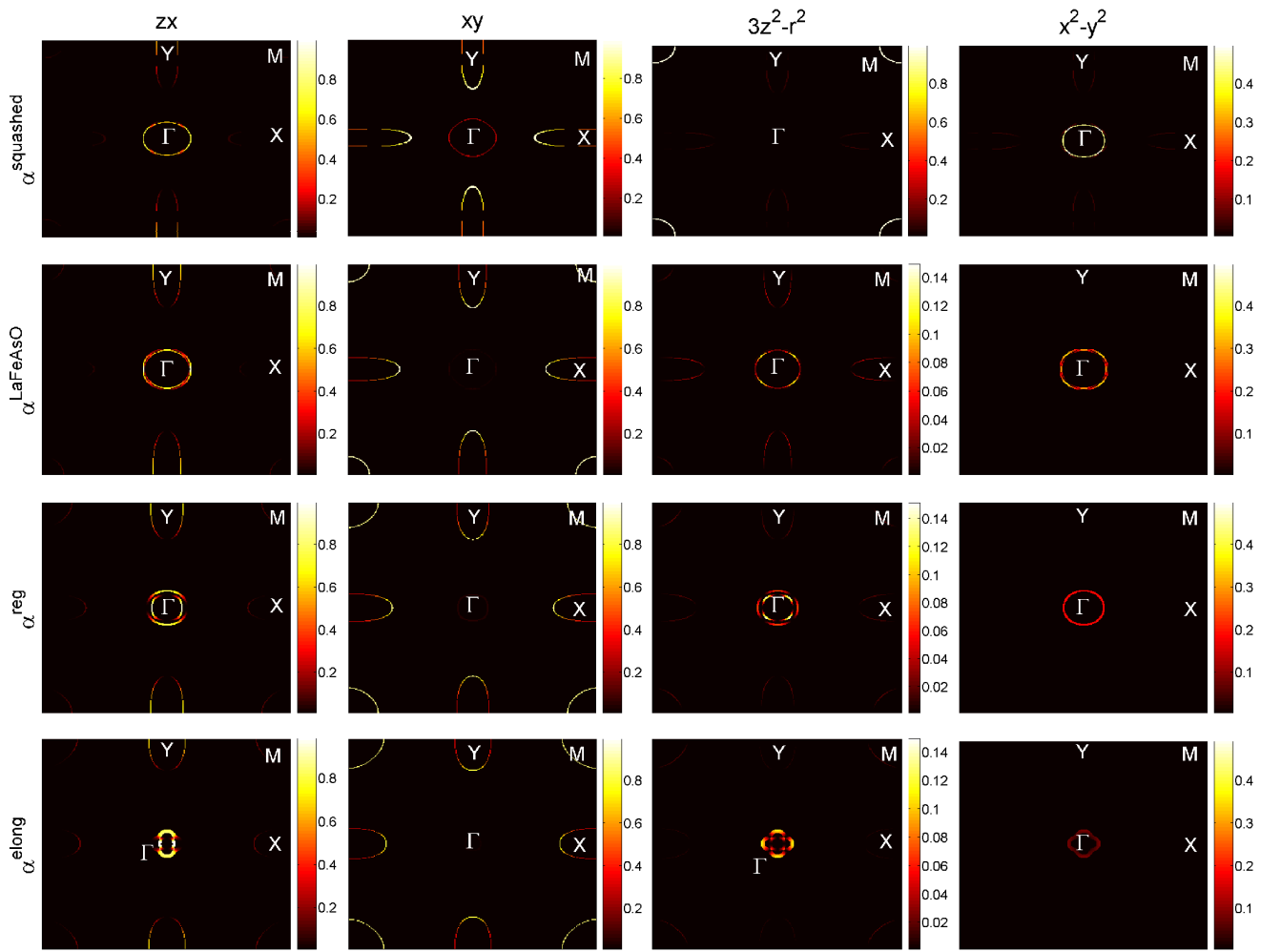}
\caption{ From left to right: Orbital content of the Fermi surface corresponding to orbitals
$zx$, $xy$, $3z^2-r^2$ and $x^2-y^2$. From top to bottom, 
each of the figures is plotted for 
$\alpha^{squashed}=29.9^o$ (as found in LaFePO), 
$\alpha^{LaFeAsO}=33.2^o$ (as found in LaFeAsO),
$\alpha^{reg}=35.3^o$ (regular tetrahedron) and $\alpha^{elong}=37,2^o$ 
(elongated tetrahedron) and the same fitting parameters used in Fig.~3. }
\label{figfsorbitalcontent}
\end{figure}

\end{widetext}

Recently it has been proposed that the anisotropic orbital makeup 
of the states on the Fermi surface is crucial to determine
 the superconducting and magnetic 
properties since it controls the anisotropy of the interband pair
scattering.    
Fig.~\ref{figfsorbitalcontent} shows that this orbital makeup is
also sensitive to changes in $\alpha$. 
Such orbital dependence on $\alpha$ 
will influence the value
and anisotropy of the pair scattering potential.  
The most dramatic example of such sensitivity 
is the change of the hole pocket at M 
from $3z^2-r^2$, for $\alpha=29.9^o$ to $xy$, character for larger $\alpha$
discussed above. 
The $zx$, $yz$ weight in the hole pockets in $\Gamma$ also reverses. 
For
$\alpha^{squashed}$, 
 the $zx$ weight is larger around $(k_F,0)$ in the inner pocket and
$(0,k_F)$ in the outer pocket. On the
contrary, for the  regular tetrahedron $\alpha^{reg}$ the situation is the 
opposite: the $zx$
weight is larger around $(0,k_F)$ in the inner pocket and $(k_F,0)$ in the
outer pocket. 
Note that, in the reference frame that we use, $zx$ and $yz$ orbitals lie
  in the plane of the Fe-Fe bonds and are not directed towards the diagonals.
Other $\alpha$-dependent effects  seen in Fig.~\ref{figfsorbitalcontent} include smaller $x^2-y^2$
weight in the hole pockets in
$\Gamma$ for larger $\alpha$
and changes in the $3z^2-r^2$ content of the hole pockets. 

The importance of nesting between electron and hole pockets has been
  emphasized in weak coupling models\cite{mazin08,raghu08,korshunov08,chubukov08,kuroki08} which place interband scattering at the
  origin of the magnetic and superconducting properties of iron pnictides.  
Due to the change in shape of electron and hole bands with $\alpha$ the amount
of nesting will be sensitive to changes in the Fe-As angle.
According to Fig.~5, for the fitting parameter used, 
the best nesting conditions are found between the inner
hole pocket in $\Gamma$ and the electron pockets in X and Y for the regular
tetrahedron case. More recently, it has been argued that the scattering
strength is not simply a consequence of nesting, but it reflects the orbital
weight structure factors. The effective pairing strength is larger between
fermions which belong to the same orbital. In Fig.~6 it can be appreciated
that this nesting is intraorbital, between segments of the inner hole pocket
and those of the electron pocket around Y with $d_{zx}$ character (and around
X with $d_{yz}$ character, not shown). Interestingly, the same result was
obtained in ab-initio studies of the evolution of the Fermi surface of
BaFe$_2$As$_2$ under pressure.

\section{Conclusions}

In conclusion, we have developed a five-orbital tight-binding model to
describe FeAs layers in iron pnictides  with hopping amplitudes calculated within the Slater-Koster framework. This
method to determine the hopping amplitudes allows to analyze the dependence of
the band structure on the Fe-As angle $\alpha$. 
A good description of the bands, including its orbital content, 
can be obtained using only four fitting
constants to parametrize all the hopping amplitudes  which compare 
well with LDA bands. 

The flexibility to study changes in the
lattice and the  small number of fitting parameters make this model
 a good starting point 
to which interactions can be added in order to
study the magnetic 
and superconducting properties.
We have  shown that changes in iron-pnictogen angle $\alpha$ induce 
changes in the shape 
of the Fermi surface and in its orbital makeup. 
In particular, in agreement with LDA calculations the hole pocket in 
$(\pi,\pi)$  ($\Gamma'$ in the reduced Brillouin zone) has $3z^2-r^2$ character for $\alpha^{squashed}$ and $xy$ character 
for $\alpha^{LaFeAsO}$. In our tight-binding model these changes can be 
understood 
in terms of the evolution of the hopping parameters with $\alpha$. 
This sensitivity extends to the nesting properties of the Fermi surface.
In a weak coupling scenario changes of the shape, nesting and orbital content 
of the Fermi surface with $\alpha$ could be at
the origin of the different superconducting order parameters, critical 
temperature and magnetic properties 
found in different iron pnictides. Within the strong coupling\cite{yildirim08,daghofer08,si08,haule08} point of view 
the superexchange 
interactions will also be affected by changes in $\alpha$ via the hopping 
amplitudes.

We have  benefited from conversations with E. Cappelluti and D.H. Lee. We
acknowledge funding from Ministerio de Ciencia e Innovaci\'on
through Grants No. FIS2005-05478-C02-01, FIS2008-00124/FIS and MAT2006-03741 
and Ram\'on y Cajal contracts, and from
Consejer\'{\i}a de Educaci\'on de la Comunidad Aut\'onoma de Madrid
and CSIC through Grants No. CCG07-CSIC/ESP-2323 and CCG08-CSIC/ESP-3518.

\appendix

\section{Hopping Amplitudes in Slater-Koster framework}
\label{app:hoppings}
In this appendix we give the expressions for the hopping amplitudes calculated
within the Slater-Koster formalism\cite{slater54}. 
Both Fe-Fe direct hopping, as well as
hopping via As are included in the expressions below. 
Fe-Fe direct hopping is described via first $(dd\sigma)_1$, $(dd\pi)_1$ and
$(dd\delta)_1$ and second  $(dd\sigma)_2$, $(dd\pi)_2$ and
$(dd\delta)_2$ nearest neighbors overlap integrals between d-orbitals. 
Fe-As hopping amplitudes are restricted to first nearest neighbors and involve 
orbital
overlap integral  $pd\sigma$ and $pd\pi$ between As-p and Fe-d orbitals, 
which mediate both first and second nearest neighbors hopping between 
Fe atoms. To compute hopping via arsenic atoms to second order in perturbation
theory, we neglect the difference among the onsite energies of the $d$ 
orbitals and that among the onsite energies of the $p$ orbitals and take them
equal to $\epsilon_d$ and $\epsilon_p$ respectively. 
It is only in the expression for the indirect hopping amplitudes that the
difference between the on-site energies of the d-Fe orbitals has been
neglected. The values $\epsilon_\alpha$ are explicitly included and taken into
account in the tight-binding expression (\ref{tightbindingreal}). 
The resulting finite hopping amplitudes are, between first nearest neighbors:
\begin{widetext}
\begin{eqnarray}
t_{xy,xy}^{x,y}&=&\frac{1}{|\epsilon_p-\epsilon_d|}\left(-{\frac{3}{2}}\, pd\sigma^2-2 \,pd\pi^2+2\sqrt{3}\, pd\sigma \, pd\pi \right) \cos^4\alpha \sin^2 \alpha  +(dd\pi)_1  \\
t_{yz,yz}^{x} &=& \frac{1}{|\epsilon_p-\epsilon_d|}\left[   \left(\frac{3}{4}
    \, pd\sigma^2 \sin^2\alpha + \sqrt{3} \,pd\sigma \, pd\pi \cos^2\alpha
  \right)  \sin^2(2\alpha) \right. \nonumber \\  & + & \left. 
    pd\pi^2 \left(\cos^2\alpha + 2 \sin^2\alpha \left[1- \cos^2\alpha (3+
        \cos(2\alpha))\right]\right)\right]
   +(dd\delta)_1\\
t_{yz,yz}^{y}&=&\frac{1}{|\epsilon_p-\epsilon_d|}\left [ \left(-\frac{3}{4} \, pd\sigma^2 + \sqrt{3}\, pd\sigma \, pd\pi\right) \sin^2(2\alpha) \sin^2\alpha - 
    pd\pi^2 \left(1 - 3 \sin^2\alpha +  \sin^2(2\alpha) \sin^2\alpha \right)\right] \nonumber \\
    &+&(dd\pi)_1 \\
t_{3z^2-r^2,3z^2-r^2}^{x,y}&=& \frac{1}{|\epsilon_p-\epsilon_d|}\left [pd\sigma^2 \sin^2 \alpha \left(\frac{1}{2}\cos^4\alpha  - \frac{1}{2} \sin^2 (2\alpha) + 
        2 \sin^4 \alpha \right)+ \frac{3}{2}\ pd\pi^2\ \cos^2 \alpha \sin^2(2
      \alpha) \right. \nonumber \\ &+& \left. \sqrt{3}\,  pd\sigma \,pd\pi
      \sin^2(2\alpha) \left(-\frac{1}{2}\cos^2\alpha  +  \sin^2 \alpha
      \right)\right ] 
        +\frac{1}{4} (dd\sigma)_1+\frac{3}{4} (dd\delta)_1 \\
t_{x^2-y^2,x^2-y^2}^{x,y}&=&\frac{3}{4} (dd\sigma)_1+\frac{1}{4} (dd\delta)_1 \\
t_{xy,yz}^{y}&=&\frac{1}{|\epsilon_p-\epsilon_d|}
\left[-\frac{3}{8\sqrt{2}} pd\sigma^2 \sin^2(2 \alpha) + \frac{\sqrt{2}}{2}\,
  pd\pi^2  \left(1 - \frac{1}{2}  \sin^2(2 \alpha) \right) +
  \frac{\sqrt{6}}{4} \,pd\sigma \, pd\pi  \sin^2 (2\alpha)\right]\sin(2\alpha) 
\\
%\end{eqnarray}
%\begin{eqnarray}
%
t_{yz,3z^2-r^2}^{x}&=& \frac{1}{|\epsilon_p-\epsilon_d|}
\left[\frac{\sqrt{3}}{4 \sqrt{2}} \, pd\sigma^2 \sin^2\alpha  \left(1-3
    \cos(2\alpha) \right)+pd\pi^2 \sqrt{\frac{3}{2}}  \left(-\frac{1}{4}+
    \cos(2\alpha)+\frac{1}{4} \cos(4\alpha) \right)  \right. \nonumber \\
&+&\left.  pd\sigma \, pd\pi \cos^2(\alpha)  \left(\sqrt{2}-\frac{3}{\sqrt{2}}
    \cos(2 \alpha) \right)\right] \sin(2\alpha) \\
t_{yz,x^2-y^2}^{x}&=&\frac{1}{|\epsilon_p-\epsilon_d|}
\left[ -\frac{\sqrt{2}}{2} pd\pi^2  \left(1-2 \cos^2\alpha \right)-\frac{\sqrt{6}}{2} \,pd\sigma \,pd\pi  \cos^2\alpha \right] \sin(2\alpha) \\
t_{3z^2-r^2,x^2-y^2}^{x}&=&\frac{1}{|\epsilon_p-\epsilon_d|} \left[
\frac{ \sqrt{3}}{2} \, pd\pi^2  \sin^2 (2\alpha) + pd\sigma \, pd\pi \cos^2
\alpha \left(1-3 \sin^2 \alpha \right) \right ]-\frac{\sqrt{3}}{4} (dd\sigma)_1+\frac{\sqrt{3}}{4} (dd\delta)_1
\end{eqnarray}
\end{widetext}
and between second nearest neighbors:
\begin{widetext}
\begin{eqnarray}
\tilde{t}_{xy,xy}&=&\frac{1}{|\epsilon_p-\epsilon_d|}
\left[- \frac{3}{4} \, pd\sigma^2 \cos^2 \alpha \cos (2 \alpha) + 
     \,pd\pi^2 \cos(2\alpha) \sin^2\alpha - \frac{\sqrt{3}}{2} \, pd\sigma \, pd\pi \, \sin^2(2\alpha)\right] \cos^2 \alpha \nonumber\\
    &+&\frac{3}{4} (dd\sigma)_2+\frac{1}{4} (dd\delta)_2 \\
\tilde{t}_{yz,yz}&=&\frac{1}{|\epsilon_p-\epsilon_d|} \left[
\left(\frac{3}{8} \, pd\sigma^2 - 
    \frac{\sqrt{3}}{2} \, pd\sigma\, pd\pi \right) \cos(2\alpha) \sin^2(2
  \alpha) + \frac{1}{4} pd\pi^2 \left(1 - \frac{5}{2} \cos(2 \alpha) -
    \frac{1}{2} \cos(6 \alpha)\right)\right]\nonumber\\ &+& \frac{1}{2} (dd\pi)_2+\frac{1}{2} (dd\delta)_2   \\
% %    
\tilde{t}_{3z^2-r^2,3z^2-r^2}&=&\frac{1}{|\epsilon_p-\epsilon_d|} \left[
pd\sigma^2 \left(-\frac{1}{4} \cos^6 \alpha + \frac{5}{4} \cos^4\alpha
  \sin^2\alpha - 2 \cos^2 \alpha \sin^4 \alpha+ \sin^6 \alpha
\right)+\frac{3}{4} pd\pi^2 \cos(2\alpha) \sin^2(2\alpha) \right. \nonumber \\
&+& \left. \frac{\sqrt{3}}{2}  pd\sigma \, pd\pi \sin^2(2\alpha) \left(3
    \sin^2\alpha-1)\right) \right ]
+ \frac{1}{4} (dd\sigma)_2+\frac{3}{4} (dd\delta)_2 \\
\tilde{t}_{x^2-y^2,x^2-y^2}&=&- \frac{1}{|\epsilon_p-\epsilon_d|}pd\pi^2 \cos^2\alpha + (dd\pi)_2 \\
\tilde{t}_{xy,yz}&=& \frac{1}{|\epsilon_p-\epsilon_d|}\frac{1}{4 \sqrt{2}} \left[3 \,pd\sigma^2 \cos^2 \alpha \cos (2 \alpha) + 
    pd\pi^2 \left(1+ \cos(4 \alpha)\right)  
    -\sqrt{3} \, pd\sigma\,  pd\pi  \left( \cos(2 \alpha)+\cos(4 \alpha)\right)\right] \sin(2\alpha) \\
\tilde{t}_{xy,3z^2-r^2}&=&   \frac{1}{|\epsilon_p-\epsilon_d|}\left [
\frac{\sqrt{3}}{8} pd\sigma^2 \cos^2 \alpha \left(\frac{3}{2} -\cos(2 \alpha)
  + \frac{3}{2} \cos(4 \alpha) \right) - \frac{\sqrt{3}}{4} pd\pi^2 \cos(2
\alpha)\sin^2(2 \alpha)\right. \nonumber \\ &+& \left.  
   \frac{1}{4} pd\sigma pd\pi \left(1+ 3 \cos(2 \alpha)  \right)\sin^2(2
   \alpha)\right ]
   -\frac{\sqrt{3}}{4} (dd\sigma)_2+\frac{\sqrt{3}}{4} (dd\delta)_2 \\
\tilde{t}_{yz,zx}&=&  \frac{1}{|\epsilon_p-\epsilon_d|}\left [  
\frac{3}{8} \, pd\sigma^2 \cos(2 \alpha) \sin^2 (2\alpha)-\frac{1}{4}\,
pd\pi^2 \left(1+ \frac{1}{2} \cos (2 \alpha) +\frac{1}{2} \cos(6 \alpha)
\right) - \frac{\sqrt{3}}{2} \, pd\sigma\, pd\pi \cos(2 \alpha) \sin^2(2
\alpha) 
\right ]\nonumber \\ &+&
\frac{1}{2} (dd\pi)_2-\frac{1}{2} (dd\delta)_2 \\
\tilde{t}_{yz,3z^2-r^2}&=& \frac{1}{|\epsilon_p-\epsilon_d|} \left[\frac{\sqrt{3}}{16}  \,pd\sigma^2  \left(3 -  
        2 \cos(2 \alpha)+ 3 \cos(4 \alpha) \right)+ \frac{\sqrt{3}}{2}
      \,pd\pi^2 \cos^2(2\alpha)  \right. \nonumber \\ &+& \left.\frac{1}{4}\, pd\sigma \, pd\pi \left(\cos(2\alpha) -3 \cos(4\alpha) \right) \right]\frac{\sin(2\alpha)}{\sqrt{2}} \\
\tilde{t}_{yz,x^2-y^2}&=& \frac{1}{|\epsilon_p-\epsilon_d|}pd\pi^2 \frac{\sin(2\alpha)}{2 \sqrt{2}}
\end{eqnarray}
\end{widetext}
Any other hopping amplitude not listed here is zero or is related by symmetry
to these ones, as discussed in Sec. II. 
The overlap integrals are treated as fitting parameters.
Due to the shorter distance between the atoms the largest contribution to hopping is expected to come
from the first nearest neighbors Fe-As 
and Fe-Fe overlap integrals. As shown in the text
it is possible to reproduce the most important features of the band structure,
including its orbital content and dependence on $\alpha$, neglecting all the 
contributions beyond these ones: $pd\sigma$, $pd\pi$, $(dd\sigma)_1$,
$(dd\pi)_1$ and $(dd\delta)_1$. 
From the above fitting we see that 
while the inclusion of direct hopping between 
Fe nearest neighbors is crucial to reproduce the band structure, direct
hopping to second Fe neighbors can be neglected. 
Giving all the energies, including 
d-d orbital overlap integrals, in units of 
$(pd\sigma)^2/|\epsilon_d-\epsilon_p|$,
and except p-d overlap integrals, which are given in units of $pd\sigma$, the
determination of the hopping amplitudes reduces to the computation of just
four fitting parameters $pd\pi$, $(dd\sigma)_1$, $(dd\pi)_1$ and 
$(dd\delta)_1$.

The formalism used here allows to study how the band
structure depends on changes in the lattice. Indirect hopping between 
Fe-atoms via As induces a dependence of the hopping
amplitudes in the angle $\alpha$ formed between the Fe-As bonds and the Fe-Fe
plane. This dependence is plotted  in Fig.~\ref{fig:parameters}. All hopping
terms show angle-dependence with the only exception of 
$t^x_{x^2-y^2,x^2-y^2}$ which does not have any indirect 
contribution. On the other hand the amplitudes which couple orbitals
$yz,zx$ with $xy,3z^2-r^2,x^2-y^2$ vanish for $\alpha=0$ when the arsenic atoms are
in the Fe-plane.

\section{Hamiltonian in momentum space}
\label{app:hamk}
In the ten-orbital reduced Brillouin zone 
$-\frac{\pi}{2}< k'_x,k'_y <\frac{\pi}{2}$ the Hamiltonian is a $10 \times 10$ 
matrix. Only the terms which mix  orbitals $zx, yz$ with $xy,3z^2-r^2,x^2-y^2$
 feel the unit-cell doubling and couple states with momentum
${\bf k'}$
with other with momentum ${\bf k'+Q}$ with ${\bf Q=(\pi,\pi)}$. Choosing the orbital basis 
$d_{{\bf k'};\alpha,\sigma},d_{{\bf
  k'+Q};\alpha,\sigma}$ in a convenient order: $\{d_{{\bf k'};yz,\sigma}, d_{{\bf
    k'};zx,\sigma},d_{{\bf k'+Q};xy,\sigma},d_{{\bf
    k'+Q};3z^2-r^2,\sigma},d_{{\bf k'+Q};x^2-y^2,\sigma}, \\ d_{{\bf k'+Q};yz,\sigma}, d_{{\bf
    k'+Q};zx,\sigma},d_{{\bf k'};xy,\sigma},d_{{\bf
    k'};3z^2-r^2,\sigma},d_{{\bf k'};x^2-y^2,\sigma}\}$ the
$ 10 \times 10$ Hamiltonian can be written as block diagonal:
\begin{displaymath}
\left (\begin{array}{c c}H_{5\times 5}({\bf k'}) & 0 \\ 0 & H_{5\times 5}({\bf
      k'+Q})\end{array}\right )
\label{matriz10por10}
\end{displaymath}
with
\begin{displaymath}
H_{5\times 5} ({\bf k'}) = \left(\begin{array} {cc} H_{2\times2} ({\bf k'})& H_{2\times
      3}({\bf k'}) \\ H_{3 \times 2} ({\bf k'})& H_{3\times 3}({\bf
      k'+Q})\end{array}\right )-(\mu-\epsilon_\gamma)\mathbb{I}
\end{displaymath}
 Here $\mathbb{I}$ is the unit matrix and $\gamma$ the orbital index in the
order given above. 
The subindices in the matrix name serve to label the matrices and indicate their dimension. 
\begin{widetext}
\begin{displaymath}
H_{2\times 2} ({\bf k'})= \left (\begin{array}{c c} 2t^y_{yz,yz}\cos k_y' +
    2t^x_{yz,yz}\cos k_x' +4 \tilde t_{yz,yz}\cos k_x'\cos k_y'& -4\tilde t_{yz,zx}
    \sin k_x'\sin k_y' \\ -4 \tilde t_{yz,zx} \sin k_x' \sin k_y' & 
2t^y_{zx,zx}\cos k_y'+2t^x_{zx,zx}\cos k_x'+4\tilde t_{zx,zx}\cos k_x' 
\cos k_y'\end{array}\right )
\end{displaymath}
\begin{displaymath}
H_{3\times 3}({\bf k'})=\left (\begin{array}{c c c}
\begin{array}{c}2t^x_{xy,xy}\left ( \cos k_x'+\cos k_y'\right ) \\ +4 \tilde t_{xy,xy}\cos k_x'
\cos k_y' \end{array}& -4 \tilde t_{xy,3z^2-r^2} \sin k_x' \sin k_y' & 0 \\   -4 \tilde
t_{xy,3z^2-r^2} \sin k_x' \sin k_y'& \begin{array}{c}2t^x_{3z^2-r^2,3z^2-r^2}\left ( \cos
  k_x'+\cos k_y'\right ) \\ +4 \tilde t_{3z^2-r^2,3z^2-r^2}\cos k_x' \cos k_y'
\end{array}&
2t^x_{x^2-y^2,3z^2-r^2}\left (\cos k_x'-\cos k_y' \right) \\0 &
  2t^x_{x^2-y^2,3z^2-r^2}\left (\cos k_x'-\cos k_y'\right )
  & \begin{array}{c}2t^x_{x^2-y^2,x^2-y^2}\left ( \cos k_x'+\cos k_y'\right ) \\ +4 \tilde t_{x^2-y^2,x^2-y^2}\cos k_x'
\cos k_y'\end{array}
\end{array}\right )
\end{displaymath}
and $H_{3 \times 2}=\left[H^*_{2 \times 3}\right ]^T$ with $H_{2\times 3}({\bf
  k'})$ equal to
\begin{displaymath}
%H_{2\times 3}({\bf k'})=
\left (
% \begin{array}{c c c} \begin{array}{c}2i\sin k_y'\left
%       (t^y_{yz,xy}\right.\\ \left. -2\tilde t_{yz,xy}
%   \cos k_x'\right )\end{array} &  \begin{array}{c}2i\sin k_x'\left
%   (t^x_{yz,3z^2-r^2}\right. \\ \left. -2\tilde t_{yz,3z^2-r^2}\cos
%     k_y'\right )\end{array}&  \begin{array}{c}2i\sin k_x'\left
%     (t^x_{yz,x^2-y^2}\right. \\ \left. -2\tilde t_{yz,x^2-y^2}\cos
%     k_y'\right) \end{array}\\  \begin{array}{c}2i\sin k_x'\left
%     (t^x_{zx,xy}\right. \\ \left. -2\tilde t_{zx,xy}\cos
%     k_y'\right ) \end{array}&  \begin{array}{c}2i\sin k_y'\left
%     (t^y_{zx,3z^2-r^2}\right. \\\left. -2\tilde t_{zx,3z^2-r^2}
%   \cos k_x'\right) \end{array}&  \begin{array}{c}2i\sin k_y'\left
%   (t^y_{zx,x^2-y^2}\right. \\ \left.- 2\tilde t_{zx,x^2-y^2}
%   \cos k_x'\right)\end{array}
% \end{array}
\begin{array}{c c c} 2i\sin k_y'\left
      (t^y_{yz,xy}\right. \left. -2\tilde t_{yz,xy}
  \cos k_x'\right ) & 2i\sin k_x'\left
  (t^x_{yz,3z^2-r^2}\right.  \left. -2\tilde t_{yz,3z^2-r^2}\cos
    k_y'\right )&  2i\sin k_x'\left
    (t^x_{yz,x^2-y^2}\right.  \left. -2\tilde t_{yz,x^2-y^2}\cos
    k_y'\right) \\  2i\sin k_x'\left
    (t^x_{zx,xy}\right. \left. -2\tilde t_{zx,xy}\cos
    k_y'\right ) &  2i\sin k_y'\left
    (t^y_{zx,3z^2-r^2}\right. \left. -2\tilde t_{zx,3z^2-r^2}
  \cos k_x'\right) &  2i\sin k_y'\left
  (t^y_{zx,x^2-y^2}\right. \left.- 2\tilde t_{zx,x^2-y^2}
  \cos k_x'\right)
\end{array}
\right )
\end{displaymath}
\end{widetext}
In the above mentioned basis order, the block diagonal form given
provides a natural way\cite{patricklee08,yu09}  to unfold the Brillouin zone
 (see Fig. \ref{fig:lattice}), that is, to
define  ${\bf k}={\bf k'}$ for orbitals $yz$ and $zx$ and  ${\bf k}={\bf k'+Q}$
 for $xy, 3z^2-r^2$ and $x^2-y^2$. In this extended Brillouin zone 
$ -\pi< k_x,k_y <\pi$ the Hamiltonian is given by $H_{5  \times 5}({\bf
  k})$. All the figures and expressions in the main text are given in the
unfolded $k$-space.

 One advantage of  the  present tight binding model is the possibility
  to understand many features of the band structure. In particular exactly at $\Gamma$
  and $M$ none of the five bands show orbital mixing and simple expressions
  follow for their energies:
\begin{equation}
E_{\gamma}=\pm 2t^y_{\gamma,\gamma} \pm 2t^x_{\gamma,\gamma}+4\tilde
t_{\gamma,\gamma}+\epsilon_\gamma-\mu  
\label{egammayz}
\end{equation}
with $\gamma=yz$,$zx$. Plus (minus) sign applies  at $\Gamma$ (M). 
\begin{equation}
E_{\gamma}=\mp 4t^x_{\gamma,\gamma} +4 \tilde
t_{\gamma,\gamma}+\epsilon_\gamma-\mu  
\label{egammaxy}
\end{equation}
for $\gamma=xy$, $3z^2-r^2$ and $x^2-y^2$. Minus (plus) sign applies  at $\Gamma$ (M).
From (\ref{egammayz}) and (\ref{equalities}) the degeneracy of $yz$ and $zx$
bands at $\Gamma$ and M follows. This two-fold degeneracy is clearly seen at
the top of the hole bands which cross the Fermi level in $\Gamma$ and in the
two highest in energy bands at $M$. As discussed in the text the dependence of
$E_{xy}(M)$ on $\alpha$ originates in the sensitivity of $\tilde t_{xy,xy}$ to
changes in the angle. The same dependence is present in $E_{xy}(\Gamma)$ which
shifts with $\alpha$ in the same way as $E_{xy}(M)$ does,  namely, decreases as
the angle is squashed, keeping $E_{xy}(\Gamma)-E_{xy}(M)=-8t^x_{xy,xy}$ almost
unchanged. On the contrary, due to the combined dependence of first and second
nearest neighbors in $\alpha$ both $E_{3z^2-r^2}(M)$ and 
$E_{3z^2-r^2}(\Gamma)-E_{3z^2-r^2}(M)$ decrease when the angle is
elongated,  the latter becoming eventually negative. Dependence of the hopping
parameters on $\alpha$ are plotted in Fig.~2.

At $X$ and $Y$ only $3z^2-r^2$ and $x^2-y^2$ mix, the energies of 
the other orbitals can be expressed in  the simple form:
\begin{equation}
E_{\gamma}(X,Y)=\pm 2t^y_{\gamma,\gamma}\mp 2t^x_{\gamma,\gamma} -4\tilde
t_{\gamma,\gamma}+\epsilon_\gamma-\mu  
\end{equation}
for $yz$ and $zx$. First sign applies for $X$  and second for $Y$. The
degeneracy found in $\Gamma$ and $M$ is broken, but their energies are related
by symmetry $E_{yz,yz}(X)=E_{zx,zx}(Y)$ and $E_{yz,yz}(Y)=E_{zx,zx}(X)$. Due
to the  sign which precedes $\tilde t_{xy,xy}$ in the expression for  
the energy corresponding to  the $xy$ orbital in 
$X$
and $Y$, 
%\begin{equation}
$E_{xy}(X,Y)=-4\tilde t_{xy,xy}+\epsilon_{xy}-\mu$,
%\end{equation}
the shift of $E_{xy}(X,Y)$ with $\alpha$ is oppossite to that found at $\Gamma$ and $M$.

In the present paper we have neglected the dependence of the crystal field
splitting on the angle $\alpha$ but, due to the simplicity of the 
expressions for the energy 
at the symmetry points,  guessing its effect in the bandstructure
is straightforward.  In particular, a possible change in crystal field of $xy$
with $\alpha$ would shift  $E_{xy}$ in the same amount in $\Gamma$, $M$, $X$ and
$Y$, contrary to the effect produced by the angle-dependent hopping parameters.

Another interesting feature regards the mixing between orbitals along 
the high
symmetry lines $\Gamma Y$, $\Gamma X$, $MX$ and $MY$. 
Along $\Gamma Y$  and $MX$ $xy/yz$ bands cross $zx/x^2-y^2/3z^2-r^2$ bands
without hybridization resulting in Dirac points. These Dirac 
points can be observed in Fig.~3 and Fig.~4. Along $MX$ there are crossings
between the two upmost bands and between the second and third bands (the later
crossing being only present for $\alpha^{squashed}$ in Fig.~4). Along $\Gamma Y$
such crossings appear between the two upmost bands and between the two lower
 ones, the former crossing being absent in the case of  $\alpha^{squashed}$. 
Along
this direction a Dirac point close to the Fermi level is also found at the
crossing between $zx$ and $xy$ derived bands, as discussed in the main text 
and mentioned previously by other authors.\cite{kuroki09-2,zhai09} The same physics 
appears along $\Gamma X$ and $MY$ with the  interchange of $yz$ for $zx$.

%
%\bibliography{hoppings-square_prb}

\begin{thebibliography}{50}
\expandafter\ifx\csname natexlab\endcsname\relax\def\natexlab#1{#1}\fi
\expandafter\ifx\csname bibnamefont\endcsname\relax
  \def\bibnamefont#1{#1}\fi
\expandafter\ifx\csname bibfnamefont\endcsname\relax
  \def\bibfnamefont#1{#1}\fi
\expandafter\ifx\csname citenamefont\endcsname\relax
  \def\citenamefont#1{#1}\fi
\expandafter\ifx\csname url\endcsname\relax
  \def\url#1{\texttt{#1}}\fi
\expandafter\ifx\csname urlprefix\endcsname\relax\def\urlprefix{URL }\fi
\providecommand{\bibinfo}[2]{#2}
\providecommand{\eprint}[2][]{\url{#2}}

\bibitem[{\citenamefont{Kamihara et~al.}(2006)\citenamefont{Kamihara,
  Hiramatsu, Hirano, Kawamura, Yanagi, Kamiya, and Hosono}}]{kamihara06}
\bibinfo{author}{\bibfnamefont{Y.}~\bibnamefont{Kamihara}},
  \bibinfo{author}{\bibfnamefont{H.}~\bibnamefont{Hiramatsu}},
  \bibinfo{author}{\bibfnamefont{M.}~\bibnamefont{Hirano}},
  \bibinfo{author}{\bibfnamefont{R.}~\bibnamefont{Kawamura}},
  \bibinfo{author}{\bibfnamefont{H.}~\bibnamefont{Yanagi}},
  \bibinfo{author}{\bibfnamefont{T.}~\bibnamefont{Kamiya}}, \bibnamefont{and}
  \bibinfo{author}{\bibfnamefont{H.}~\bibnamefont{Hosono}},
  \bibinfo{journal}{J. Am. Chem. Soc} \textbf{\bibinfo{volume}{128}},
  \bibinfo{pages}{10013} (\bibinfo{year}{2006}).

\bibitem[{\citenamefont{Kamihara et~al.}(2008)\citenamefont{Kamihara, Watanabe,
  Hirano, and Hosono}}]{kamihara08}
\bibinfo{author}{\bibfnamefont{Y.}~\bibnamefont{Kamihara}},
  \bibinfo{author}{\bibfnamefont{T.}~\bibnamefont{Watanabe}},
  \bibinfo{author}{\bibfnamefont{M.}~\bibnamefont{Hirano}}, \bibnamefont{and}
  \bibinfo{author}{\bibfnamefont{H.}~\bibnamefont{Hosono}},
  \bibinfo{journal}{J. Am. Chem. Soc} \textbf{\bibinfo{volume}{130}},
  \bibinfo{pages}{3296} (\bibinfo{year}{2008}).

\bibitem[{\citenamefont{Nomura et~al.}(2009)\citenamefont{Nomura, Inoue,
  Matsuishi, Hirano, Kim, Kato, Takata, and Hosono}}]{nomura09}
\bibinfo{author}{\bibfnamefont{T.}~\bibnamefont{Nomura}},
  \bibinfo{author}{\bibfnamefont{Y.}~\bibnamefont{Inoue}},
  \bibinfo{author}{\bibfnamefont{S.}~\bibnamefont{Matsuishi}},
  \bibinfo{author}{\bibfnamefont{M.}~\bibnamefont{Hirano}},
  \bibinfo{author}{\bibfnamefont{J.~E.} \bibnamefont{Kim}},
  \bibinfo{author}{\bibfnamefont{K.}~\bibnamefont{Kato}},
  \bibinfo{author}{\bibfnamefont{M.}~\bibnamefont{Takata}}, \bibnamefont{and}
  \bibinfo{author}{\bibfnamefont{H.}~\bibnamefont{Hosono}},
  \bibinfo{journal}{Supercond. Sci. Technol.} \textbf{\bibinfo{volume}{22}},
  \bibinfo{pages}{055016} (\bibinfo{year}{2009}).

\bibitem[{\citenamefont{Ogino et~al.}(2009)\citenamefont{Ogino, Matsumura,
  Katsura, Ushiyama, Horii, Kishio, and Shimoyama}}]{ogino09}
\bibinfo{author}{\bibfnamefont{H.}~\bibnamefont{Ogino}},
  \bibinfo{author}{\bibfnamefont{Y.}~\bibnamefont{Matsumura}},
  \bibinfo{author}{\bibfnamefont{Y.}~\bibnamefont{Katsura}},
  \bibinfo{author}{\bibfnamefont{K.}~\bibnamefont{Ushiyama}},
  \bibinfo{author}{\bibfnamefont{S.}~\bibnamefont{Horii}},
  \bibinfo{author}{\bibfnamefont{K.}~\bibnamefont{Kishio}}, \bibnamefont{and}
  \bibinfo{author}{\bibfnamefont{J.}~\bibnamefont{Shimoyama}},
  \bibinfo{journal}{Supercond. Sci. Technol.} \textbf{\bibinfo{volume}{22}},
  \bibinfo{pages}{085001} (\bibinfo{year}{2009}).

\bibitem[{\citenamefont{Rotter et~al.}(2008)\citenamefont{Rotter, Tegel, and
  Johrendt}}]{rotter08}
\bibinfo{author}{\bibfnamefont{M.}~\bibnamefont{Rotter}},
  \bibinfo{author}{\bibfnamefont{M.}~\bibnamefont{Tegel}}, \bibnamefont{and}
  \bibinfo{author}{\bibfnamefont{D.}~\bibnamefont{Johrendt}},
  \bibinfo{journal}{Phys. Rev. Lett.} \textbf{\bibinfo{volume}{101}},
  \bibinfo{pages}{107006} (\bibinfo{year}{2008}).

\bibitem[{\citenamefont{Zhao et~al.}(2008)\citenamefont{Zhao, Huang, de~la
  Cruz, Li, Lynn, Chen, Green, Chen, Li, Li et~al.}}]{zhao08}
\bibinfo{author}{\bibfnamefont{J.}~\bibnamefont{Zhao}},
  \bibinfo{author}{\bibfnamefont{Q.}~\bibnamefont{Huang}},
  \bibinfo{author}{\bibfnamefont{C.}~\bibnamefont{de~la Cruz}},
  \bibinfo{author}{\bibfnamefont{S.}~\bibnamefont{Li}},
  \bibinfo{author}{\bibfnamefont{J.~W.} \bibnamefont{Lynn}},
  \bibinfo{author}{\bibfnamefont{Y.}~\bibnamefont{Chen}},
  \bibinfo{author}{\bibfnamefont{M.~A.} \bibnamefont{Green}},
  \bibinfo{author}{\bibfnamefont{G.~F.} \bibnamefont{Chen}},
  \bibinfo{author}{\bibfnamefont{G.}~\bibnamefont{Li}},
  \bibinfo{author}{\bibfnamefont{Z.}~\bibnamefont{Li}}, \bibnamefont{et~al.},
  \bibinfo{journal}{Nature Materials} \textbf{\bibinfo{volume}{7}},
  \bibinfo{pages}{953} (\bibinfo{year}{2008}).

\bibitem[{\citenamefont{Kimber et~al.}(2009)\citenamefont{Kimber, Kreyssig,
  Zhang, Jeschke, Valenti, Yokaichiya, Colombier, Yan, Hansen, Chatterji
  et~al.}}]{kimber09}
\bibinfo{author}{\bibfnamefont{S.~A.~J.} \bibnamefont{Kimber}},
  \bibinfo{author}{\bibfnamefont{A.}~\bibnamefont{Kreyssig}},
  \bibinfo{author}{\bibfnamefont{Y.-Z.} \bibnamefont{Zhang}},
  \bibinfo{author}{\bibfnamefont{H.~O.} \bibnamefont{Jeschke}},
  \bibinfo{author}{\bibfnamefont{R.}~\bibnamefont{Valenti}},
  \bibinfo{author}{\bibfnamefont{F.}~\bibnamefont{Yokaichiya}},
  \bibinfo{author}{\bibfnamefont{E.}~\bibnamefont{Colombier}},
  \bibinfo{author}{\bibfnamefont{J.}~\bibnamefont{Yan}},
  \bibinfo{author}{\bibfnamefont{T.~C.} \bibnamefont{Hansen}},
  \bibinfo{author}{\bibfnamefont{T.}~\bibnamefont{Chatterji}},
  \bibnamefont{et~al.}, \bibinfo{journal}{Nature Materials}
  \textbf{\bibinfo{volume}{8}}, \bibinfo{pages}{471} (\bibinfo{year}{2009}).

\bibitem[{\citenamefont{Lee et~al.}(2008)\citenamefont{Lee, Iyo, Eisaki, Kito,
  Fernandez-Diaz, Ito, Kihou, Matsuhata, Braden, and Yamada}}]{Lee08}
\bibinfo{author}{\bibfnamefont{C.~H.} \bibnamefont{Lee}},
  \bibinfo{author}{\bibfnamefont{A.}~\bibnamefont{Iyo}},
  \bibinfo{author}{\bibfnamefont{H.}~\bibnamefont{Eisaki}},
  \bibinfo{author}{\bibfnamefont{H.}~\bibnamefont{Kito}},
  \bibinfo{author}{\bibfnamefont{M.~T.} \bibnamefont{Fernandez-Diaz}},
  \bibinfo{author}{\bibfnamefont{T.}~\bibnamefont{Ito}},
  \bibinfo{author}{\bibfnamefont{K.}~\bibnamefont{Kihou}},
  \bibinfo{author}{\bibfnamefont{H.}~\bibnamefont{Matsuhata}},
  \bibinfo{author}{\bibfnamefont{M.}~\bibnamefont{Braden}}, \bibnamefont{and}
  \bibinfo{author}{\bibfnamefont{K.}~\bibnamefont{Yamada}},
  \bibinfo{journal}{J. Phys. Soc. Jpn.} \textbf{\bibinfo{volume}{77}},
  \bibinfo{pages}{083704} (\bibinfo{year}{2008}).

\bibitem[{\citenamefont{McQueen et~al.}(2008)\citenamefont{McQueen, Regulacio,
  Williams, Huang, Lynn, Hor, West, Green, and Cava}}]{mcqueen08}
\bibinfo{author}{\bibfnamefont{T.~M.} \bibnamefont{McQueen}},
  \bibinfo{author}{\bibfnamefont{M.}~\bibnamefont{Regulacio}},
  \bibinfo{author}{\bibfnamefont{A.~J.} \bibnamefont{Williams}},
  \bibinfo{author}{\bibfnamefont{Q.}~\bibnamefont{Huang}},
  \bibinfo{author}{\bibfnamefont{J.~W.} \bibnamefont{Lynn}},
  \bibinfo{author}{\bibfnamefont{Y.~S.} \bibnamefont{Hor}},
  \bibinfo{author}{\bibfnamefont{D.~V.} \bibnamefont{West}},
  \bibinfo{author}{\bibfnamefont{M.~A.} \bibnamefont{Green}}, \bibnamefont{and}
  \bibinfo{author}{\bibfnamefont{R.~J.} \bibnamefont{Cava}},
  \bibinfo{journal}{Phys. Rev. B} \textbf{\bibinfo{volume}{78}},
  \bibinfo{pages}{024521} (\bibinfo{year}{2008}).

\bibitem[{\citenamefont{Kuroki et~al.}(2009)\citenamefont{Kuroki, Usui, Onari,
  Arita, and Aoki}}]{kuroki09-2}
\bibinfo{author}{\bibfnamefont{K.}~\bibnamefont{Kuroki}},
  \bibinfo{author}{\bibfnamefont{H.}~\bibnamefont{Usui}},
  \bibinfo{author}{\bibfnamefont{S.}~\bibnamefont{Onari}},
  \bibinfo{author}{\bibfnamefont{R.}~\bibnamefont{Arita}}, \bibnamefont{and}
  \bibinfo{author}{\bibfnamefont{H.}~\bibnamefont{Aoki}},
  \bibinfo{journal}{Phys. Rev. B} \textbf{\bibinfo{volume}{79}},
  \bibinfo{pages}{224511} (\bibinfo{year}{2009}).

\bibitem[{\citenamefont{Leb\`egue}(2007)}]{lebegue07}
\bibinfo{author}{\bibfnamefont{S.}~\bibnamefont{Leb\`egue}},
  \bibinfo{journal}{Phys. Rev. B} \textbf{\bibinfo{volume}{75}},
  \bibinfo{pages}{035110} (\bibinfo{year}{2007}).

\bibitem[{\citenamefont{Singh and Du}(2008)}]{singh08}
\bibinfo{author}{\bibfnamefont{D.~J.} \bibnamefont{Singh}} \bibnamefont{and}
  \bibinfo{author}{\bibfnamefont{M.-H.} \bibnamefont{Du}},
  \bibinfo{journal}{Phys. Rev. Lett.} \textbf{\bibinfo{volume}{100}},
  \bibinfo{pages}{237003} (\bibinfo{year}{2008}).

\bibitem[{\citenamefont{Mazin et~al.}(2008{\natexlab{a}})\citenamefont{Mazin,
  Johannes, Boeri, and Koepernik}}]{mazin08-2}
\bibinfo{author}{\bibfnamefont{I.~I.} \bibnamefont{Mazin}},
  \bibinfo{author}{\bibfnamefont{M.~D.} \bibnamefont{Johannes}},
  \bibinfo{author}{\bibfnamefont{L.}~\bibnamefont{Boeri}}, \bibnamefont{and}
  \bibinfo{author}{\bibfnamefont{K.}~\bibnamefont{Koepernik}},
  \bibinfo{journal}{Phys. Rev. B} \textbf{\bibinfo{volume}{78}},
  \bibinfo{pages}{085104} (\bibinfo{year}{2008}{\natexlab{a}}).

\bibitem[{\citenamefont{Mazin et~al.}(2008{\natexlab{b}})\citenamefont{Mazin,
  Singh, Johannes, and Du}}]{mazin08}
\bibinfo{author}{\bibfnamefont{I.~I.} \bibnamefont{Mazin}},
  \bibinfo{author}{\bibfnamefont{D.~J.} \bibnamefont{Singh}},
  \bibinfo{author}{\bibfnamefont{M.~D.} \bibnamefont{Johannes}},
  \bibnamefont{and} \bibinfo{author}{\bibfnamefont{M.~H.} \bibnamefont{Du}},
  \bibinfo{journal}{Phys. Rev. Lett.} \textbf{\bibinfo{volume}{101}},
  \bibinfo{pages}{057003} (\bibinfo{year}{2008}{\natexlab{b}}).

\bibitem[{\citenamefont{Coldea et~al.}(2008)\citenamefont{Coldea, Fletcher,
  Carrington, Analytis, Bangura, Chu, Erickson, Fisher, Hussey, and
  McDonald}}]{coldea08}
\bibinfo{author}{\bibfnamefont{A.~I.} \bibnamefont{Coldea}},
  \bibinfo{author}{\bibfnamefont{J.~D.} \bibnamefont{Fletcher}},
  \bibinfo{author}{\bibfnamefont{A.}~\bibnamefont{Carrington}},
  \bibinfo{author}{\bibfnamefont{J.~G.} \bibnamefont{Analytis}},
  \bibinfo{author}{\bibfnamefont{A.~F.} \bibnamefont{Bangura}},
  \bibinfo{author}{\bibfnamefont{J.~H.} \bibnamefont{Chu}},
  \bibinfo{author}{\bibfnamefont{A.~S.} \bibnamefont{Erickson}},
  \bibinfo{author}{\bibfnamefont{I.~R.} \bibnamefont{Fisher}},
  \bibinfo{author}{\bibfnamefont{N.~E.} \bibnamefont{Hussey}},
  \bibnamefont{and} \bibinfo{author}{\bibfnamefont{R.~D.}
  \bibnamefont{McDonald}}, \bibinfo{journal}{Phys. Rev. Lett.}
  \textbf{\bibinfo{volume}{101}}, \bibinfo{pages}{216402}
  (\bibinfo{year}{2008}).

\bibitem[{\citenamefont{Liu et~al.}(2008{\natexlab{a}})\citenamefont{Liu,
  Samolyuk, Lee, Ni, Kondo, Santander-Syro, Bud'ko, McChesney, Rotenberg, Valla
  et~al.}}]{liu08-2}
\bibinfo{author}{\bibfnamefont{C.}~\bibnamefont{Liu}},
  \bibinfo{author}{\bibfnamefont{G.~D.} \bibnamefont{Samolyuk}},
  \bibinfo{author}{\bibfnamefont{Y.}~\bibnamefont{Lee}},
  \bibinfo{author}{\bibfnamefont{N.}~\bibnamefont{Ni}},
  \bibinfo{author}{\bibfnamefont{T.}~\bibnamefont{Kondo}},
  \bibinfo{author}{\bibfnamefont{A.~F.} \bibnamefont{Santander-Syro}},
  \bibinfo{author}{\bibfnamefont{S.~L.} \bibnamefont{Bud'ko}},
  \bibinfo{author}{\bibfnamefont{J.~L.} \bibnamefont{McChesney}},
  \bibinfo{author}{\bibfnamefont{E.}~\bibnamefont{Rotenberg}},
  \bibinfo{author}{\bibfnamefont{T.}~\bibnamefont{Valla}},
  \bibnamefont{et~al.}, \bibinfo{journal}{Phys. Rev. Lett.}
  \textbf{\bibinfo{volume}{101}}, \bibinfo{pages}{177005}
  (\bibinfo{year}{2008}{\natexlab{a}}).

\bibitem[{\citenamefont{Liu et~al.}(2008{\natexlab{b}})\citenamefont{Liu,
  Zhang, Zhao, Jia, Meng, Liu, Dong, Chen, Luo, Wang et~al.}}]{liu08-zhou}
\bibinfo{author}{\bibfnamefont{H.}~\bibnamefont{Liu}},
  \bibinfo{author}{\bibfnamefont{W.}~\bibnamefont{Zhang}},
  \bibinfo{author}{\bibfnamefont{L.}~\bibnamefont{Zhao}},
  \bibinfo{author}{\bibfnamefont{X.}~\bibnamefont{Jia}},
  \bibinfo{author}{\bibfnamefont{J.}~\bibnamefont{Meng}},
  \bibinfo{author}{\bibfnamefont{G.}~\bibnamefont{Liu}},
  \bibinfo{author}{\bibfnamefont{X.}~\bibnamefont{Dong}},
  \bibinfo{author}{\bibfnamefont{G.~F.} \bibnamefont{Chen}},
  \bibinfo{author}{\bibfnamefont{J.~L.} \bibnamefont{Luo}},
  \bibinfo{author}{\bibfnamefont{N.~L.} \bibnamefont{Wang}},
  \bibnamefont{et~al.}, \bibinfo{journal}{Phys. Rev. B}
  \textbf{\bibinfo{volume}{78}}, \bibinfo{pages}{1845}
  (\bibinfo{year}{2008}{\natexlab{b}}).

\bibitem[{\citenamefont{Lu et~al.}(2008)\citenamefont{Lu, Yi, Mo, Erickson,
  Analytis, Chu, Singh, Hussain, Geballe, Fisher et~al.}}]{lu08}
\bibinfo{author}{\bibfnamefont{D.~H.} \bibnamefont{Lu}},
  \bibinfo{author}{\bibfnamefont{M.}~\bibnamefont{Yi}},
  \bibinfo{author}{\bibfnamefont{S.-K.} \bibnamefont{Mo}},
  \bibinfo{author}{\bibfnamefont{A.~S.} \bibnamefont{Erickson}},
  \bibinfo{author}{\bibfnamefont{J.}~\bibnamefont{Analytis}},
  \bibinfo{author}{\bibfnamefont{J.-H.} \bibnamefont{Chu}},
  \bibinfo{author}{\bibfnamefont{D.~J.} \bibnamefont{Singh}},
  \bibinfo{author}{\bibfnamefont{Z.}~\bibnamefont{Hussain}},
  \bibinfo{author}{\bibfnamefont{T.~H.} \bibnamefont{Geballe}},
  \bibinfo{author}{\bibfnamefont{I.~R.} \bibnamefont{Fisher}},
  \bibnamefont{et~al.}, \bibinfo{journal}{Nature}
  \textbf{\bibinfo{volume}{455}}, \bibinfo{pages}{81} (\bibinfo{year}{2008}).

\bibitem[{\citenamefont{Ding et~al.}(2008)\citenamefont{Ding, Nakayama,
  Richard, Souma, Sato, Takahashi, Neupane, Xu, Pan, Federov
  et~al.}}]{ding08-2}
\bibinfo{author}{\bibfnamefont{H.}~\bibnamefont{Ding}},
  \bibinfo{author}{\bibfnamefont{K.}~\bibnamefont{Nakayama}},
  \bibinfo{author}{\bibfnamefont{P.}~\bibnamefont{Richard}},
  \bibinfo{author}{\bibfnamefont{S.}~\bibnamefont{Souma}},
  \bibinfo{author}{\bibfnamefont{T.}~\bibnamefont{Sato}},
  \bibinfo{author}{\bibfnamefont{T.}~\bibnamefont{Takahashi}},
  \bibinfo{author}{\bibfnamefont{M.}~\bibnamefont{Neupane}},
  \bibinfo{author}{\bibfnamefont{Y.-M.} \bibnamefont{Xu}},
  \bibinfo{author}{\bibfnamefont{Z.-H.} \bibnamefont{Pan}},
  \bibinfo{author}{\bibfnamefont{A.~V.} \bibnamefont{Federov}},
  \bibnamefont{et~al.} (\bibinfo{year}{2008}), \bibinfo{note}{arXiv:0812.0534}.

\bibitem[{\citenamefont{Zabolotnyy et~al.}(2009)\citenamefont{Zabolotnyy,
  Inosov, Evtushinsky, Koitzsch, Kordyuk, Sun, Park, Haug, Hinkov, Boris
  et~al.}}]{zabolotnyy08}
\bibinfo{author}{\bibfnamefont{V.~B.} \bibnamefont{Zabolotnyy}},
  \bibinfo{author}{\bibfnamefont{D.~S.} \bibnamefont{Inosov}},
  \bibinfo{author}{\bibfnamefont{D.~V.} \bibnamefont{Evtushinsky}},
  \bibinfo{author}{\bibfnamefont{A.}~\bibnamefont{Koitzsch}},
  \bibinfo{author}{\bibfnamefont{A.~A.} \bibnamefont{Kordyuk}},
  \bibinfo{author}{\bibfnamefont{G.~L.} \bibnamefont{Sun}},
  \bibinfo{author}{\bibfnamefont{J.~T.} \bibnamefont{Park}},
  \bibinfo{author}{\bibfnamefont{D.}~\bibnamefont{Haug}},
  \bibinfo{author}{\bibfnamefont{V.}~\bibnamefont{Hinkov}},
  \bibinfo{author}{\bibfnamefont{A.~V.} \bibnamefont{Boris}},
  \bibnamefont{et~al.}, \bibinfo{journal}{Nature}
  \textbf{\bibinfo{volume}{457}}, \bibinfo{pages}{569} (\bibinfo{year}{2009}).

\bibitem[{\citenamefont{Yao et~al.}(2009)\citenamefont{Yao, Li, and
  Wang}}]{yao09}
\bibinfo{author}{\bibfnamefont{Z.-J.} \bibnamefont{Yao}},
  \bibinfo{author}{\bibfnamefont{J.-X.} \bibnamefont{Li}}, \bibnamefont{and}
  \bibinfo{author}{\bibfnamefont{Z.~D.} \bibnamefont{Wang}},
  \bibinfo{journal}{New.J. of Physics} \textbf{\bibinfo{volume}{11}},
  \bibinfo{pages}{025009} (\bibinfo{year}{2009}).

\bibitem[{\citenamefont{Raghu et~al.}(2008)\citenamefont{Raghu, Qi, Liu,
  Scalapino, and Zhang}}]{raghu08}
\bibinfo{author}{\bibfnamefont{S.}~\bibnamefont{Raghu}},
  \bibinfo{author}{\bibfnamefont{X.~L.} \bibnamefont{Qi}},
  \bibinfo{author}{\bibfnamefont{C.-X.} \bibnamefont{Liu}},
  \bibinfo{author}{\bibfnamefont{D.~J.} \bibnamefont{Scalapino}},
  \bibnamefont{and} \bibinfo{author}{\bibfnamefont{S.-C.} \bibnamefont{Zhang}},
  \bibinfo{journal}{Phys. Rev. B} \textbf{\bibinfo{volume}{77}},
  \bibinfo{pages}{220503} (\bibinfo{year}{2008}).

\bibitem[{\citenamefont{Korshunov and Eremin}(2008)}]{korshunov08}
\bibinfo{author}{\bibfnamefont{M.~M.} \bibnamefont{Korshunov}}
  \bibnamefont{and} \bibinfo{author}{\bibfnamefont{I.}~\bibnamefont{Eremin}},
  \bibinfo{journal}{Phys. Rev. B} \textbf{\bibinfo{volume}{78}},
  \bibinfo{pages}{140509(R)} (\bibinfo{year}{2008}).

\bibitem[{\citenamefont{Maier et~al.}(2009)\citenamefont{Maier, Graser,
  Scalapino, and Hirschfeld}}]{maier09}
\bibinfo{author}{\bibfnamefont{T.~A.} \bibnamefont{Maier}},
  \bibinfo{author}{\bibfnamefont{S.}~\bibnamefont{Graser}},
  \bibinfo{author}{\bibfnamefont{D.~J.} \bibnamefont{Scalapino}},
  \bibnamefont{and} \bibinfo{author}{\bibfnamefont{P.~J.}
  \bibnamefont{Hirschfeld}}, \bibinfo{journal}{Phys. Rev. B}
  \textbf{\bibinfo{volume}{79}}, \bibinfo{pages}{224510}
  (\bibinfo{year}{2009}).

\bibitem[{\citenamefont{Zhai et~al.}(2009)\citenamefont{Zhai, Wang, and
  Lee}}]{zhai09}
\bibinfo{author}{\bibfnamefont{H.}~\bibnamefont{Zhai}},
  \bibinfo{author}{\bibfnamefont{F.}~\bibnamefont{Wang}}, \bibnamefont{and}
  \bibinfo{author}{\bibfnamefont{D.-H.} \bibnamefont{Lee}}
  (\bibinfo{year}{2009}), \bibinfo{note}{arXiv:0905.1711v1}.

\bibitem[{\citenamefont{Boeri et~al.}(2008)\citenamefont{Boeri, Dolgov, and
  Golubov}}]{boeri08}
\bibinfo{author}{\bibfnamefont{L.}~\bibnamefont{Boeri}},
  \bibinfo{author}{\bibfnamefont{O.~V.} \bibnamefont{Dolgov}},
  \bibnamefont{and} \bibinfo{author}{\bibfnamefont{A.~A.}
  \bibnamefont{Golubov}}, \bibinfo{journal}{Phys. Rev. Lett.}
  \textbf{\bibinfo{volume}{101}}, \bibinfo{pages}{085104}
  (\bibinfo{year}{2008}).

\bibitem[{\citenamefont{Vildosola et~al.}(2008)\citenamefont{Vildosola,
  Pourovskii, Arita, Biermann, and Georges}}]{vildosola08}
\bibinfo{author}{\bibfnamefont{V.}~\bibnamefont{Vildosola}},
  \bibinfo{author}{\bibfnamefont{L.}~\bibnamefont{Pourovskii}},
  \bibinfo{author}{\bibfnamefont{R.}~\bibnamefont{Arita}},
  \bibinfo{author}{\bibfnamefont{S.}~\bibnamefont{Biermann}}, \bibnamefont{and}
  \bibinfo{author}{\bibfnamefont{A.}~\bibnamefont{Georges}},
  \bibinfo{journal}{Phys. Rev. B} \textbf{\bibinfo{volume}{130}},
  \bibinfo{pages}{064518} (\bibinfo{year}{2008}).

\bibitem[{\citenamefont{Leb\`egue et~al.}(2009)\citenamefont{Leb\`egue, Yin,
  and Pickett}}]{lebegue09}
\bibinfo{author}{\bibfnamefont{S.}~\bibnamefont{Leb\`egue}},
  \bibinfo{author}{\bibfnamefont{Z.~P.} \bibnamefont{Yin}}, \bibnamefont{and}
  \bibinfo{author}{\bibfnamefont{W.~E.} \bibnamefont{Pickett}},
  \bibinfo{journal}{New J. of Physics} \textbf{\bibinfo{volume}{11}},
  \bibinfo{pages}{025004} (\bibinfo{year}{2009}).

\bibitem[{\citenamefont{Hsieh et~al.}(2008)\citenamefont{Hsieh, Xia, Wray,
  Qian, Gomes, Yazdani, Chen, Luo, Wang, and Hasan}}]{hsieh08}
\bibinfo{author}{\bibfnamefont{D.}~\bibnamefont{Hsieh}},
  \bibinfo{author}{\bibfnamefont{Y.}~\bibnamefont{Xia}},
  \bibinfo{author}{\bibfnamefont{L.}~\bibnamefont{Wray}},
  \bibinfo{author}{\bibfnamefont{D.}~\bibnamefont{Qian}},
  \bibinfo{author}{\bibfnamefont{K.~K.} \bibnamefont{Gomes}},
  \bibinfo{author}{\bibfnamefont{A.}~\bibnamefont{Yazdani}},
  \bibinfo{author}{\bibfnamefont{G.~F.} \bibnamefont{Chen}},
  \bibinfo{author}{\bibfnamefont{J.~L.} \bibnamefont{Luo}},
  \bibinfo{author}{\bibfnamefont{N.~L.} \bibnamefont{Wang}}, \bibnamefont{and}
  \bibinfo{author}{\bibfnamefont{M.~Z.} \bibnamefont{Hasan}}
  (\bibinfo{year}{2008}), \bibinfo{note}{arXiv:0812.2289}.

\bibitem[{\citenamefont{Fink et~al.}(2008)\citenamefont{Fink, Thirupathaiah,
  Ovsyannikov, D\"urr, Follath, Huang, de~Jong, Golden, Zhang, Jescke
  et~al.}}]{fink09}
\bibinfo{author}{\bibfnamefont{J.}~\bibnamefont{Fink}},
  \bibinfo{author}{\bibfnamefont{S.}~\bibnamefont{Thirupathaiah}},
  \bibinfo{author}{\bibfnamefont{R.}~\bibnamefont{Ovsyannikov}},
  \bibinfo{author}{\bibfnamefont{H.~A.} \bibnamefont{D\"urr}},
  \bibinfo{author}{\bibfnamefont{R.}~\bibnamefont{Follath}},
  \bibinfo{author}{\bibfnamefont{Y.}~\bibnamefont{Huang}},
  \bibinfo{author}{\bibfnamefont{S.}~\bibnamefont{de~Jong}},
  \bibinfo{author}{\bibfnamefont{M.~S.} \bibnamefont{Golden}},
  \bibinfo{author}{\bibfnamefont{Y.-Z.} \bibnamefont{Zhang}},
  \bibinfo{author}{\bibfnamefont{H.~O.} \bibnamefont{Jescke}},
  \bibnamefont{et~al.}, \bibinfo{journal}{Phys. Rev. B}
  \textbf{\bibinfo{volume}{79}}, \bibinfo{pages}{155118}
  (\bibinfo{year}{2008}).

\bibitem[{\citenamefont{Shimojima et~al.}(2009)\citenamefont{Shimojima,
  Ishizaka, Ishida, Katayama, Ogushi, Kiss, Okawa, Togashi, Wang, Chen
  et~al.}}]{shimojima09}
\bibinfo{author}{\bibfnamefont{T.}~\bibnamefont{Shimojima}},
  \bibinfo{author}{\bibfnamefont{K.}~\bibnamefont{Ishizaka}},
  \bibinfo{author}{\bibfnamefont{Y.}~\bibnamefont{Ishida}},
  \bibinfo{author}{\bibfnamefont{N.}~\bibnamefont{Katayama}},
  \bibinfo{author}{\bibfnamefont{K.}~\bibnamefont{Ogushi}},
  \bibinfo{author}{\bibfnamefont{T.}~\bibnamefont{Kiss}},
  \bibinfo{author}{\bibfnamefont{M.}~\bibnamefont{Okawa}},
  \bibinfo{author}{\bibfnamefont{T.}~\bibnamefont{Togashi}},
  \bibinfo{author}{\bibfnamefont{X.-Y.} \bibnamefont{Wang}},
  \bibinfo{author}{\bibfnamefont{C.-T.} \bibnamefont{Chen}},
  \bibnamefont{et~al.} (\bibinfo{year}{2009}), \bibinfo{note}{arXiv:0904.1632}.

\bibitem[{\citenamefont{Zhang et~al.}(2009)\citenamefont{Zhang, Zhou, Chen,
  Wei, Xu, Yang, Fang, Tsai, Cao, Xu et~al.}}]{zhang09}
\bibinfo{author}{\bibfnamefont{Y.}~\bibnamefont{Zhang}},
  \bibinfo{author}{\bibfnamefont{B.}~\bibnamefont{Zhou}},
  \bibinfo{author}{\bibfnamefont{F.}~\bibnamefont{Chen}},
  \bibinfo{author}{\bibfnamefont{J.}~\bibnamefont{Wei}},
  \bibinfo{author}{\bibfnamefont{M.}~\bibnamefont{Xu}},
  \bibinfo{author}{\bibfnamefont{L.~X.} \bibnamefont{Yang}},
  \bibinfo{author}{\bibfnamefont{C.}~\bibnamefont{Fang}},
  \bibinfo{author}{\bibfnamefont{W.~F.} \bibnamefont{Tsai}},
  \bibinfo{author}{\bibfnamefont{G.~H.} \bibnamefont{Cao}},
  \bibinfo{author}{\bibfnamefont{Z.~A.} \bibnamefont{Xu}}, \bibnamefont{et~al.}
  (\bibinfo{year}{2009}), \bibinfo{note}{arXiv:0904.4022}.

\bibitem[{\citenamefont{Daghofer et~al.}(2008)\citenamefont{Daghofer, Moreo,
  Riera, Arrigoni, and Dagotto}}]{daghofer08}
\bibinfo{author}{\bibfnamefont{M.}~\bibnamefont{Daghofer}},
  \bibinfo{author}{\bibfnamefont{A.}~\bibnamefont{Moreo}},
  \bibinfo{author}{\bibfnamefont{J.~A.} \bibnamefont{Riera}},
  \bibinfo{author}{\bibfnamefont{E.}~\bibnamefont{Arrigoni}}, \bibnamefont{and}
  \bibinfo{author}{\bibfnamefont{E.}~\bibnamefont{Dagotto}},
  \bibinfo{journal}{Phys. Rev. Lett.} \textbf{\bibinfo{volume}{101}},
  \bibinfo{pages}{237004} (\bibinfo{year}{2008}).

\bibitem[{\citenamefont{Stanescu et~al.}(2008)\citenamefont{Stanescu, Galitski,
  and Sarma}}]{stanescu08}
\bibinfo{author}{\bibfnamefont{T.~D.} \bibnamefont{Stanescu}},
  \bibinfo{author}{\bibfnamefont{V.}~\bibnamefont{Galitski}}, \bibnamefont{and}
  \bibinfo{author}{\bibfnamefont{S.~D.} \bibnamefont{Sarma}},
  \bibinfo{journal}{Phys. Rev. B} \textbf{\bibinfo{volume}{78}},
  \bibinfo{pages}{195114} (\bibinfo{year}{2008}).

\bibitem[{\citenamefont{Lee and Wen}(2008)}]{patricklee08}
\bibinfo{author}{\bibfnamefont{P.~A.} \bibnamefont{Lee}} \bibnamefont{and}
  \bibinfo{author}{\bibfnamefont{X.-G.} \bibnamefont{Wen}},
  \bibinfo{journal}{Phys. Rev. B} \textbf{\bibinfo{volume}{78}},
  \bibinfo{pages}{144517} (\bibinfo{year}{2008}).

\bibitem[{\citenamefont{Kuroki et~al.}(2008)\citenamefont{Kuroki, Onari, Arita,
  Usui, Tanaka, Kontani, and Aoki}}]{kuroki08}
\bibinfo{author}{\bibfnamefont{K.}~\bibnamefont{Kuroki}},
  \bibinfo{author}{\bibfnamefont{S.}~\bibnamefont{Onari}},
  \bibinfo{author}{\bibfnamefont{R.}~\bibnamefont{Arita}},
  \bibinfo{author}{\bibfnamefont{H.}~\bibnamefont{Usui}},
  \bibinfo{author}{\bibfnamefont{Y.}~\bibnamefont{Tanaka}},
  \bibinfo{author}{\bibfnamefont{H.}~\bibnamefont{Kontani}}, \bibnamefont{and}
  \bibinfo{author}{\bibfnamefont{H.}~\bibnamefont{Aoki}},
  \bibinfo{journal}{Phys. Rev. Lett.} \textbf{\bibinfo{volume}{101}},
  \bibinfo{pages}{087004} (\bibinfo{year}{2008}).

\bibitem[{\citenamefont{Cvetkovic and Tesanovic}(2009)}]{cvetkovic09}
\bibinfo{author}{\bibfnamefont{V.}~\bibnamefont{Cvetkovic}} \bibnamefont{and}
  \bibinfo{author}{\bibfnamefont{Z.}~\bibnamefont{Tesanovic}},
  \bibinfo{journal}{Europhys. Lett.} \textbf{\bibinfo{volume}{85}},
  \bibinfo{pages}{37002} (\bibinfo{year}{2009}).

\bibitem[{\citenamefont{Eschrig and Koepernik}(2009)}]{eschrig09}
\bibinfo{author}{\bibfnamefont{H.}~\bibnamefont{Eschrig}} \bibnamefont{and}
  \bibinfo{author}{\bibfnamefont{K.}~\bibnamefont{Koepernik}}
  (\bibinfo{year}{2009}), \bibinfo{note}{arXiv:0905.4844v2}.

\bibitem[{\citenamefont{Slater and Koster}(1954)}]{slater54}
\bibinfo{author}{\bibfnamefont{J.~C.} \bibnamefont{Slater}} \bibnamefont{and}
  \bibinfo{author}{\bibfnamefont{G.~F.} \bibnamefont{Koster}},
  \bibinfo{journal}{Phys. Rev.} \textbf{\bibinfo{volume}{94}},
  \bibinfo{pages}{1498} (\bibinfo{year}{1954}).

\bibitem[{\citenamefont{Yu et~al.}(2009)\citenamefont{Yu, Trinh, Moreo,
  Daghofer, Riera, Haas, and Dagotto}}]{yu09}
\bibinfo{author}{\bibfnamefont{R.}~\bibnamefont{Yu}},
  \bibinfo{author}{\bibfnamefont{K.~T.} \bibnamefont{Trinh}},
  \bibinfo{author}{\bibfnamefont{A.}~\bibnamefont{Moreo}},
  \bibinfo{author}{\bibfnamefont{M.}~\bibnamefont{Daghofer}},
  \bibinfo{author}{\bibfnamefont{J.~A.} \bibnamefont{Riera}},
  \bibinfo{author}{\bibfnamefont{S.}~\bibnamefont{Haas}}, \bibnamefont{and}
  \bibinfo{author}{\bibfnamefont{E.}~\bibnamefont{Dagotto}},
  \bibinfo{journal}{Phys. Rev. B} \textbf{\bibinfo{volume}{79}},
  \bibinfo{pages}{104510} (\bibinfo{year}{2009}).

\bibitem[{\citenamefont{Graser et~al.}(2009)\citenamefont{Graser, Maier,
  Hirschfeld, and Scalapino}}]{graser09}
\bibinfo{author}{\bibfnamefont{S.}~\bibnamefont{Graser}},
  \bibinfo{author}{\bibfnamefont{T.~A.} \bibnamefont{Maier}},
  \bibinfo{author}{\bibfnamefont{P.~J.} \bibnamefont{Hirschfeld}},
  \bibnamefont{and} \bibinfo{author}{\bibfnamefont{D.~J.}
  \bibnamefont{Scalapino}}, \bibinfo{journal}{New J. Phys.}
  \textbf{\bibinfo{volume}{11}}, \bibinfo{pages}{025016}
  (\bibinfo{year}{2009}).

\bibitem[{\citenamefont{Calder\'on et~al.}(2009)\citenamefont{Calder\'on,
  Valenzuela, and Bascones}}]{calderon09}
\bibinfo{author}{\bibfnamefont{M.~J.} \bibnamefont{Calder\'on}},
  \bibinfo{author}{\bibfnamefont{B.}~\bibnamefont{Valenzuela}},
  \bibnamefont{and} \bibinfo{author}{\bibfnamefont{E.}~\bibnamefont{Bascones}},
  \bibinfo{journal}{New J. Phys.} \textbf{\bibinfo{volume}{11}},
  \bibinfo{pages}{013051} (\bibinfo{year}{2009}).

\bibitem[{\citenamefont{Moreo et~al.}(2009)\citenamefont{Moreo, Daghofer,
  Riera, and Dagotto}}]{moreo09}
\bibinfo{author}{\bibfnamefont{A.}~\bibnamefont{Moreo}},
  \bibinfo{author}{\bibfnamefont{M.}~\bibnamefont{Daghofer}},
  \bibinfo{author}{\bibfnamefont{J.~A.} \bibnamefont{Riera}}, \bibnamefont{and}
  \bibinfo{author}{\bibfnamefont{E.}~\bibnamefont{Dagotto}},
  \bibinfo{journal}{Phys. Rev. B} \textbf{\bibinfo{volume}{79}},
  \bibinfo{pages}{134502} (\bibinfo{year}{2009}).

\bibitem[{\citenamefont{Chubukov et~al.}(2008)\citenamefont{Chubukov, Efremov,
  and Eremin}}]{chubukov08}
\bibinfo{author}{\bibfnamefont{A.~V.} \bibnamefont{Chubukov}},
  \bibinfo{author}{\bibfnamefont{D.}~\bibnamefont{Efremov}}, \bibnamefont{and}
  \bibinfo{author}{\bibfnamefont{I.}~\bibnamefont{Eremin}},
  \bibinfo{journal}{Phys. Rev. B} \textbf{\bibinfo{volume}{78}},
  \bibinfo{pages}{134512} (\bibinfo{year}{2008}).

\bibitem[{\citenamefont{Yildirim}(2008)}]{yildirim08}
\bibinfo{author}{\bibfnamefont{T.}~\bibnamefont{Yildirim}},
  \bibinfo{journal}{Physical Review Letters} \textbf{\bibinfo{volume}{101}},
  \bibinfo{pages}{057010} (\bibinfo{year}{2008}).

\bibitem[{\citenamefont{Si and Abrahams}(2008)}]{si08}
\bibinfo{author}{\bibfnamefont{Q.}~\bibnamefont{Si}} \bibnamefont{and}
  \bibinfo{author}{\bibfnamefont{E.}~\bibnamefont{Abrahams}},
  \bibinfo{journal}{Phys. Rev. Lett.} \textbf{\bibinfo{volume}{101}},
  \bibinfo{pages}{076401} (\bibinfo{year}{2008}).

\bibitem[{\citenamefont{Haule et~al.}(2008)\citenamefont{Haule, Shim, and
  Kotliar}}]{haule08}
\bibinfo{author}{\bibfnamefont{K.}~\bibnamefont{Haule}},
  \bibinfo{author}{\bibfnamefont{J.~H.} \bibnamefont{Shim}}, \bibnamefont{and}
  \bibinfo{author}{\bibfnamefont{G.}~\bibnamefont{Kotliar}},
  \bibinfo{journal}{Phys. Rev. Lett.} \textbf{\bibinfo{volume}{101}},
  \bibinfo{pages}{226402} (\bibinfo{year}{2008}).

\bibitem[{\citenamefont{de~la Cruz et~al.}(2008)\citenamefont{de~la Cruz,
  Q.-Huang, Lynn, Li, Ratcliff, Zaretsky, Mook, Chen, Luo, Wang
  et~al.}}]{delacruz08}
\bibinfo{author}{\bibfnamefont{C.}~\bibnamefont{de~la Cruz}},
  \bibinfo{author}{\bibnamefont{Q.-Huang}},
  \bibinfo{author}{\bibfnamefont{J.~W.} \bibnamefont{Lynn}},
  \bibinfo{author}{\bibfnamefont{J.}~\bibnamefont{Li}},
  \bibinfo{author}{\bibfnamefont{W.}~\bibnamefont{Ratcliff}},
  \bibinfo{author}{\bibfnamefont{J.~L.} \bibnamefont{Zaretsky}},
  \bibinfo{author}{\bibfnamefont{H.~A.} \bibnamefont{Mook}},
  \bibinfo{author}{\bibfnamefont{G.~F.} \bibnamefont{Chen}},
  \bibinfo{author}{\bibfnamefont{J.~L.} \bibnamefont{Luo}},
  \bibinfo{author}{\bibfnamefont{N.~L.} \bibnamefont{Wang}},
  \bibnamefont{et~al.}, \bibinfo{journal}{Nature}
  \textbf{\bibinfo{volume}{453}}, \bibinfo{pages}{899} (\bibinfo{year}{2008}).

\bibitem[{\citenamefont{Haule and Kotliar}(2009)}]{haule09}
\bibinfo{author}{\bibfnamefont{K.}~\bibnamefont{Haule}} \bibnamefont{and}
  \bibinfo{author}{\bibfnamefont{G.}~\bibnamefont{Kotliar}},
  \bibinfo{journal}{New J. of Phys.} \textbf{\bibinfo{volume}{11}},
  \bibinfo{pages}{025021} (\bibinfo{year}{2009}).

\bibitem[{\citenamefont{Opahle et~al.}(2009)\citenamefont{Opahle, Kandpal,
  Zhang, Gros, and Valent\`i}}]{opahle09}
\bibinfo{author}{\bibfnamefont{I.}~\bibnamefont{Opahle}},
  \bibinfo{author}{\bibfnamefont{H.~C.} \bibnamefont{Kandpal}},
  \bibinfo{author}{\bibfnamefont{Y.}~\bibnamefont{Zhang}},
  \bibinfo{author}{\bibfnamefont{C.}~\bibnamefont{Gros}}, \bibnamefont{and}
  \bibinfo{author}{\bibfnamefont{R.}~\bibnamefont{Valent\`i}},
  \bibinfo{journal}{Phys. Rev. B} \textbf{\bibinfo{volume}{79}},
  \bibinfo{pages}{024509} (\bibinfo{year}{2009}).

\end{thebibliography}

\end{document}